\documentclass[10pt,preprint2]{aastex}
\usepackage[]{graphicx}

\newcommand{\ap}{\approx}
\newcommand{\kms}{\hbox{${\rm km\;s}^{-1}$}}
\newcommand{\kmskpc}{\hbox{$\mathrm{km}\;\mathrm{s}^{-1}\;\mathrm{kpc}^{-1}$}}
\newcommand{\ha}{H$\alpha$}
\newcommand{\hi}{\ion{H}{1}}
\newcommand{\oiii}{[O~{\small III}]$\,\lambda5006.8$}  
\newcommand{\vi}{\hbox{$V\!-\!I$}}

\newcommand{\vs}{$v_\star$}
\newcommand{\vg}{$v_g$}
\newcommand{\sigstar}{$\sigma_\star$}
\newcommand{\siggas}{$\sigma_g$}
\newcommand{\htre}{$h_{3}$}
\newcommand{\hqua}{$h_{4}$}
\newcommand{\vms}{\hbox{$V_{\rm max}/\sigma$}}
\newcommand{\vmsstar}{\hbox{$(V_{\rm max}/\sigma)^{\star}$}}

\newcommand{\emax}{\hbox{$\epsilon_{max}$}}

\newcommand{\Msun}{M_{\sun}}

\slugcomment{To appear in \textit{The Astrophysical Journal}, 10 
November 2003}

\shorttitle{Inner Disks Masquerading as Bulges}
\shortauthors{Erwin et al.}

\begin{document}

\title{When Is a Bulge Not a Bulge?  Inner Disks Masquerading as
Bulges in NGC~2787 and NGC~3945}

\author{Peter Erwin, Juan Carlos Vega Beltr\'an\altaffilmark{1}}
\affil{Instituto de Astrof\'{\i}sica de Canarias, C/ Via L\'{a}ctea s/n, 
38200 La Laguna, Tenerife, Spain}
\email{erwin@ll.iac.es, jvega@ll.iac.es}
\and
\author{Alister W. Graham}
\affil{Department of Astronomy, University of Florida,
P.O.\ Box 112055, Gainesville, FL 32611, USA}
\email{graham@astro.ufl.edu}
\and
\author{John E. Beckman}
\affil{Instituto de Astrof\'{\i}sica de Canarias, C/ Via L\'{a}ctea s/n, 
38200 La Laguna, Tenerife, Spain}
\email{jeb@ll.iac.es}

% Extra affiliations:
\altaffiltext{1}{Guest investigator of the UK Astronomy Data Centre}

% The Abstract:
\begin{abstract}
We present a detailed morphological, photometric, and kinematic
analysis of two barred S0 galaxies with large, luminous \textit{inner
disks} inside their bars.  We show that these structures, in addition
to being geometrically disk-like, have exponential profiles (scale
lengths $\sim 300$--500 pc) distinct from the central, non-exponential
bulges.  We also find them to be kinematically disk-like.  The inner
disk in NGC~2787 has a luminosity roughly twice that of the bulge; but
in NGC~3945, the inner disk is almost ten times more luminous than the
bulge, which itself is extremely small (half-light radius $\approx
100$ pc, in a galaxy with an outer ring of radius $\approx 14$ kpc)
and only $\sim 5$\% of the total luminosity --- a bulge/total ratio
much more typical of an Sc galaxy.  \textit{We estimate that at least
20\% of (barred) S0 galaxies may have similar structures, which means
that their bulge/disk ratios may be significantly overestimated.}
These inner disks dominate the central light of their galaxies; they
are at least an order of magnitude larger than typical ``nuclear
disks'' found in ellipticals and early-type spirals.  Consequently,
they must affect the dynamics of the bars in which they reside.
\end{abstract}

% Keywords:
\keywords{galaxies: structure --- galaxies: disks --- galaxies: 
bulges --- galaxies: individual (NGC~2787, NGC~3945) --- galaxies: 
kinematics and dynamics}

\section{Introduction} %[x]

The presence of stellar disks around the nuclei of early-type galaxies
has attracted growing interest in recent years \citep[e.g.][and
references therein]{scorza98}.  The complexity of the central regions
of these galaxies was nicely demonstrated in the survey by
\citet{rest01}, who used WFPC2 images from the \textit{Hubble Space
Telescope} (\textit{HST}) to disentangle the detailed morphology of 67
E and S0 galaxies, finding embedded stellar disks around the nucleus
in over half their sample.  Unambiguous interpretation of detail
observed in two dimensions is often difficult, however; this was
illustrated by Rest et al., who suggested that as over half of the
circumnuclear ``disks'' detected in their survey appear misaligned
with the main galaxy, they are more probably in fact circumnuclear
bars.

This highlights a general problem: since many reported nuclear disks
have been observed in an essentially edge-on configuration, it is
easier to show that they are thin than it is to show that they are
round.  There is thus considerable scope for morphological confusion. 
This is the case, for example, in the study by \citet{seifert96}, who
found a significant number of S0 galaxies with an inner disk as well
as an outer disk.  The inner disk was revealed by subtracting off the
bulge component, and showed up clearly as having a photometric profile
with a steeper slope than that of the larger outer disk.  However,
because the galaxies studied by Seifert \& Scorza were all edge-on,
the disk-bar ambiguity alluded to above cannot be readily tackled.

In a study of moderately inclined or face-on barred S0--Sa galaxies,
\citet{erwin-sparke02} found numerous inner structures residing inside
the bars.  In the S0 galaxies, they found --- with roughly equal
frequency --- both inner bars and what they termed ``inner disks.'' 
The latter was an ad-hoc, provisional classification: elliptical (in
projection) stellar structures aligned with the outer disk, apparently
distinct from the innermost bulge or nucleus.  Some of these inner
disks could plausibly be inner bars with chance alignments;
nonetheless, the large number of such structures, their larger sizes
when compared to the inner bars, and the fact that they were
restricted almost entirely to the S0 galaxies, all suggested a
population distinct from inner bars.  This reinforces the ambiguity
noted above --- early-type galaxies can have inner disks \textit{and}
inner bars, which may not be easily distinguished when seen edge-on. 
But because the Erwin \& Sparke galaxies are \textit{not} edge-on, it
is somewhat easier to determine the geometry of these structures.

Some of these putative inner disks are quite large --- well over a
kiloparsec in size, or up to $\sim 20$\% of their galaxy's
25th-magnitude radius ($R_{25}$).  In addition, some are as large as
$\sim 40$\% of the size of the bars they were found inside, which
raises some intriguing dynamical questions: how did such large inner
structures form, and how can they coexist with the bars?

An equally significant issue is this: given that early-type disk
galaxies can have distinct inner disks, how often might inner disks in
the more face-on galaxies --- where we cannot easily identify the
(spheroidal) bulge by the fact that it protrudes above and below the
disk --- be mistaken for \textit{bulges}?  If, as is often the case,
bulges are only defined photometrically (as an inner excess in the
surface brightness profile over the extrapolated outer disk profile),
then a high surface-brightness inner disk, distinct from the outer
disk, could be (mis)classified as the galaxy's bulge.  Such a
``pseudobulge'' \citep{kormendy01,kormendy-texas} is a plausible
result of bar-driven gas inflow, followed by star formation
\citep{kormendy93}; the result would be a high-surface-brightness disk
(not necessarily exponential) inside the main disk.

This issue was raised over twenty years ago by
\citet{kormendy82a,kormendy82b}, who studied the bulge kinematics of
both barred and unbarred \citep{kormendy-ill83} galaxies.  While the
unbarred bulges --- and some of the barred bulges --- had kinematics
consistent with the classical model of bulges as small, rotationally
supported ellipticals, several of the bulges in barred galaxies had
kinematics dominated by rotation.  Thus, kinematically at least, some
``bulges'' appeared to be more like disks.  Morphological evidence
that some bulges are disk-like was discussed by \citet{kormendy93}. 
This includes cases where disk structures (e.g., spirals, bars, and
star formation) are clearly visible within the (photometric) bulge
region; the set of examples has been expanded by \textit{HST} imaging
of spiral galaxy centers \citep[e.g.,][]{carollo97,carollo98}. 
Morphologies such as this indicate that the inner regions can have
significant, even dominant, disk components, though they do not
necessarily exclude a classical bulge component as well.  Kormendy 
also pointed out that the quasi-2D bulge-disk decompositions of 
\citet{kent85,kent87} include a number of galaxies where the derived 
ellipticity of the ``bulge'' component is equal to or greater than 
that of the disk, suggesting that the inner light is dominated by a 
disk or flattened triaxial structure (e.g., a small bar).

In this paper we explore these issues by examining in detail,
photometrically and also kinematically, two of the inner-disk galaxies
studied by \citet{erwin-sparke02,erwin-sparke03}.  These objects ---
the SB0 galaxies NGC~2787 and NGC~3945 --- are two of the strongest
candidates for having large, distinct inner disks.  We use a
combination of ground-based and \textit{HST} images to test how
disk-like their inner structures actually are.  We find that they are
plausible \textit{geometric} disks: that is, they are consistent with
approximately round, flat structures seen at the same inclination as
the rest of the galaxy.  We also find that they have smooth
exponential radial light profiles, similar to those of many outer
disks of spiral and lenticular galaxies.  In addition, they are
clearly distinct from the smaller, rounder and non-exponential bulges
at their centers.  Finally, we find at least some evidence that they
are \textit{kinematic} stellar disks: they are dynamically cool, and
probably dominated by circular motions.  In this, we are following
\citet{kormendy82a,kormendy93} who drew attention to the relatively
fast rotation and low velocity dispersion in the ``bulge'' region
(within 20\arcsec{} of the center) of NGC~3945, arguing that this
region was kinematically much more like a disk than a bulge.

The existence of disks of this sort --- and confirmation that they are
distinct from central bulges --- points to a problem for identifying
and measuring bulges in early-type barred galaxies: what fraction of
the classical bulge is really a disk, i.e. a planar and kinematically
cool component, rather than a spheroidal, hot component?  We show that
for these two galaxies this problem is quite real.  In galaxies with
this composite morphology, the classical decomposition of the galaxy
as a whole into an outer disk and an inner bulge can give very
misleading results.  Nonetheless, the existence of inner disks helps
to solve the problem that they themselves create, since at least in
some cases the ``true'' spheroidal bulge component --- rounder,
kinematically hot and, in these cases, non-exponential --- can be
cleanly separated from an inner disk, using unequivocal photometric
criteria.  By effecting such a separation for the present galaxies we
show that past analyses have dramatically overestimated their bulge
sizes.  For NGC~3945 we find that the true bulge is astonishingly
small: its half-light radius is just over 100 pc, in a galaxy with a
bar of radial length 5 kpc, and an outer ring of radius 14 kpc; the
bulge accounts for only $\sim 5$\% of the total luminosity, a figure
more typical of Sc galaxies than of S0 galaxies.  In NGC~2787, the
bulge is larger (half-light radius $\ap 150$ pc and $\sim 10$\% of the
total luminosity), but the presence of the inner disk produces a
radial surface-brightness profile which can lead one to overestimate
the bulge size by factors of 2--5.

\section{An Overview of the Galaxies} %[x]
\label{sec:overview}

\begin{figure}
\begin{center}
\includegraphics[scale=0.9]{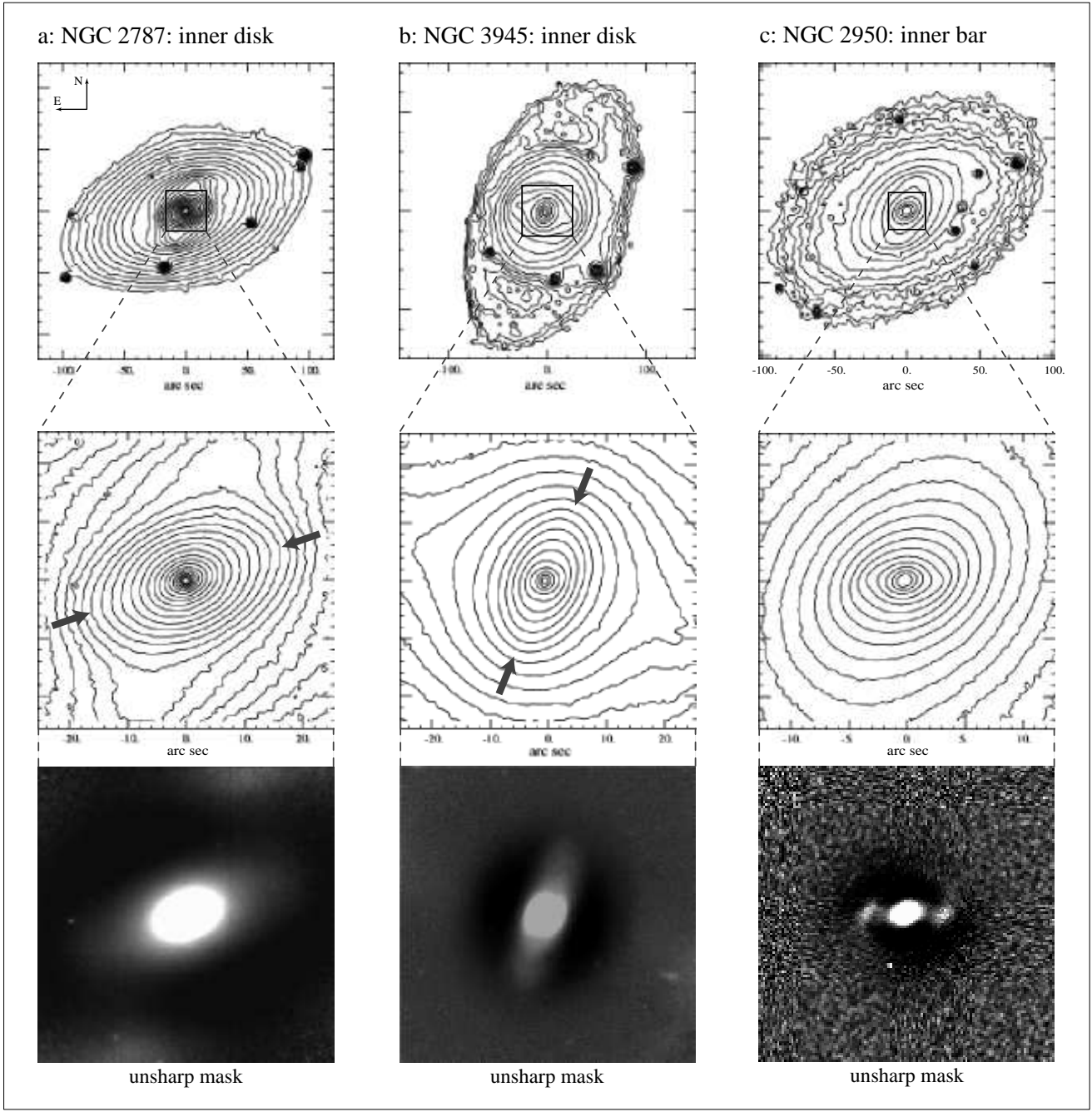}
%\epsscale{0.9}
%\plotone{f1.eps}
\end{center}

\caption{Isophotes and unsharp masks for the two inner-disk galaxies,
NGC~2787 (a) and NGC~3945 (b).  $R$-band isophotes are shown in the
top panels and in the middle panel for NGC~3945; we show $H$-band
isophotes for NGC~2787 in the middle panel.  The inner disks are
indicated by the arrows in the close-ups (middle panels), and can also
be seen in the unsharp masks (bottom panels).  For comparison, we also
show $R$-band isophotes for a double-barred galaxy \citep[NGC~2950,
from][]{erwin-sparke02}.  The inner disks of NGC~2787 and NGC~3945
appear as smooth, elliptical features in the unsharp masks, indicating
they are structurally different from inner bars such as NGC~2950's,
which shows a characteristic double-lobed feature in the unsharp mask
(see Section~\ref{sec:decomp}).\label{fig:umasks}}

\end{figure}

NGC~2787 and NGC~3945 are nearby SB0 galaxies; both are listed in
\citet{buta94} as ``representative'' of Hubble type S$0^{+}$.  General
parameters for each are given in Table~\ref{tab:general}, and
isophotal maps from our $R$-band images can be seen in
Figure~\ref{fig:umasks}.  For NGC~2787, a surface-brightness fluctuation
distance is available \citep{tonry01}; for NGC~3945, we use the
redshift with a correction for Virgo-centric infall from the
Lyon-Meudon Extragalactic Database
(LEDA)\footnote{http://leda.univ-lyon1.fr} and assume $H_{0} = 75$
\kms{} Mpc$^{-1}$.  The morphology of NGC~2787 is discussed in more
detail in \citet{erwin-sparke03}, while NGC~3945 was analyzed in
\citet{erwin99}.  We summarize some basic characteristics of each
here.

NGC~2787 has been mapped in \hi{} by \citet{shostak87}, who found a
ring with major axis $a = 6.4\arcmin$ (larger than the visible stellar
disk).  Because the position angle (140\arcdeg) and apparent
inclination (assuming the ring is intrinsically round) are different
from that of the optical disk, it suggests that this outer neutral gas
is misaligned with the stellar disk.  A large bar embedded in a lens
lies within the disk; the inner disk is inside the bar.  Much further
in, the LINER nucleus has been shown to have broad \ha{} emission
\citep{ho97d}, and \citet{sarzi01} found evidence for a $\sim 10^{8}$
solar-mass black hole, using \textit{HST} spectroscopy.  Sarzi et al.\
also pointed out the spectacular tilted dust rings in the central
regions (see their Figure~3 and Figure~2 of
\nocite{erwin-sparke02}Erwin \& Sparke 2002).  This latter feature,
combined with the \hi{} misalignment, prompted
\citet{erwin-sparke02,erwin-sparke03} to classify this galaxy as
having off-plane gas in the central regions.  Recent emission-line
spectroscopy at several position angles confirms this; we discuss this
briefly below \citep[see also][]{vega-beltran03}.

NGC~3945 is about five times more luminous than NGC~2787, and almost
five times larger in radius as well.\footnote{Comparing $D_{25}$ in
kpc.}  There is no clear outer disk in NGC~3945; instead, there is,
from the outside in, a bright, slightly asymmetric outer ring, a lens
\citep[noted by][]{kormendy79}, bounded by a dusty inner ring, and
then the outer bar.  The lens appears perpendicular to the bar; this
is probably a projection effect, since lenses are almost always
elongated parallel to bars.  Inside the outer bar is the inner disk,
which has been noted at least as far back as \citet{kormendy79}; it
has variously been described as a ``secondary bar''
\citep{kormendy79,w95} and a ``triaxial bulge'' \citep{kormendy82a}. 
\citet{erwin99} used \textit{HST} images to show that there was a
stellar nuclear ring inside the inner disk, and a probable secondary
bar inside that ring.

\section{Data}

\subsection{Imaging} %[x]

The primary imaging data are ground-based $B$ and $R$ images obtained
with the 3.5 m WIYN Telescope in Tucson, Arizona, and archival
\textit{HST} WFPC2 images in the F450W, F555W, and F814W filters
(corresponding to $B$, $V$ and $I$, respectively).\footnote{The WIYN
Observatory is a joint facility of the University of
Wisconsin-Madison, Indiana University, Yale University, and the
National Optical Astronomy Observatories.} The WIYN images were
obtained under non-photometric conditions and are not calibrated; see
\citet{erwin-sparke03} for more details of these observations.  We
\textit{do} calibrate the \textit{HST} data, in part because our
measurements of bulge and inner-disk luminosities are based on the
\textit{HST} images.  For NGC~2787, we also obtained an $H$-band image
during service/queue time (11 February 2003) with the INGRID imager on
the 4.2m William Herschel Telescope in La Palma, Spain; this is a
$1024^{2}$ near-IR imager with 0.242\arcsec{} pixels.

The WFPC2 images were processed by the standard, ``on-the-fly''
archival pipeline, then coadded with the \textsc{iraf} task
\texttt{crrej}, part of the \textsc{stsdas} package (inspection of
stars and galactic nuclei in the Planetary Camera [PC] images showed
alignment of separate exposures to within $\sim 0.1$ pixels, so the
images were combined without shifting).  Sky subtraction was done
using median pixel values in the parts of the Wide Field (WF) chips
least contaminated by galaxy light.

The photometric calibration of the PC images follows the
recommendations in \citet{holtzman95}, as used by, e.g.,
\citet{stiavelli01}.  We converted (sky-subtracted) counts per second
per pixel in the F555W and F814W images to instrumental magnitudes per
square arc second $\mu_{\rm F555W}$ and $\mu_{\rm F814W}$; these were
then transformed to surface brightness in $V$ and $I$ using the
following equations (based on Eqn.~8 in Holtzman et al.):
%     PREPRINT (two-column layout):
\begin{eqnarray}
 \mu_V  \, & = & \,  \mu_{\rm F555W} \,+\, 21.825 \,-\, 0.052(\vi) \nonumber\\
  &  & +\; 0.027(\vi)^{2}  \,+\, 2.5 \log {\rm GR} \\
 \mu_I  \, & = & \,  \mu_{\rm F814W} \,+\, 20.939 
 \,-\, 0.062(\vi) \nonumber\\
  &  & +\;  0.025(\vi)^{2} \,+\, 2.5 \log {\rm GR}.
\end{eqnarray}
%     SUMBITTED:
% \begin{equation}
%  \mu_V \,=\, \mu_{\rm F555W} \,+\, 21.825 
%  \,-\, 0.052(\vi) \,+\, 0.027(\vi)^{2} \,+\, 2.5 \log {\rm GR}
% \end{equation}
% \begin{equation}
%  \mu_I \,=\, \mu_{\rm F814W} \,+\, 20.939 
%  \,-\, 0.062(\vi) \,+\, 0.025(\vi)^{2} \,+\, 2.5 \log {\rm GR}.
% \end{equation}
The various parameters are taken from Table~7 in Holtzman et al., with
an offset of 0.1 magnitudes added to the zero points to convert to
infinite aperture (see Holtzman et al.); the gain ratio (GR) used was
that of the PC chip (1.986).  Using an initial guess for $\vi = 1.0$,
we iterated until the resulting $\vi$ values differed by less than
0.01 magnitudes; as Stiavelli et al.\ found, this generally takes only
three iterations.  Similar calibrations were made for F450W with
respect to F555W (using parameters from Table~10 of Holtzman et al.\
for the F450W filter).

We made corrections for Galactic extinction using the $A_{B}$ values
from Table~\ref{tab:general}, assuming that $A_{V} = 0.75 A_{B}$, and
$A_{I} = 0.44 A_{B}$ (with F450W, F555W, and F814W equivalent to $B$,
$V$, and $I$, respectively).

% Extinction corrections:
% Assuming A_V = 0.75 A_B, A_R = 0.56 A_B, A_I = 0.44 A_B
%    NGC 2787: A_B = 0.57 --> A_V = 0.43, A_R = 0.32, A_I = 0.25
%    NGC 3945: A_B = 0.12 --> A_V = 0.09, A_R = 0.07, A_I = 0.05

\subsection{Spectroscopy} %[x]

\begin{figure}
\begin{center}
\includegraphics[scale=0.95]{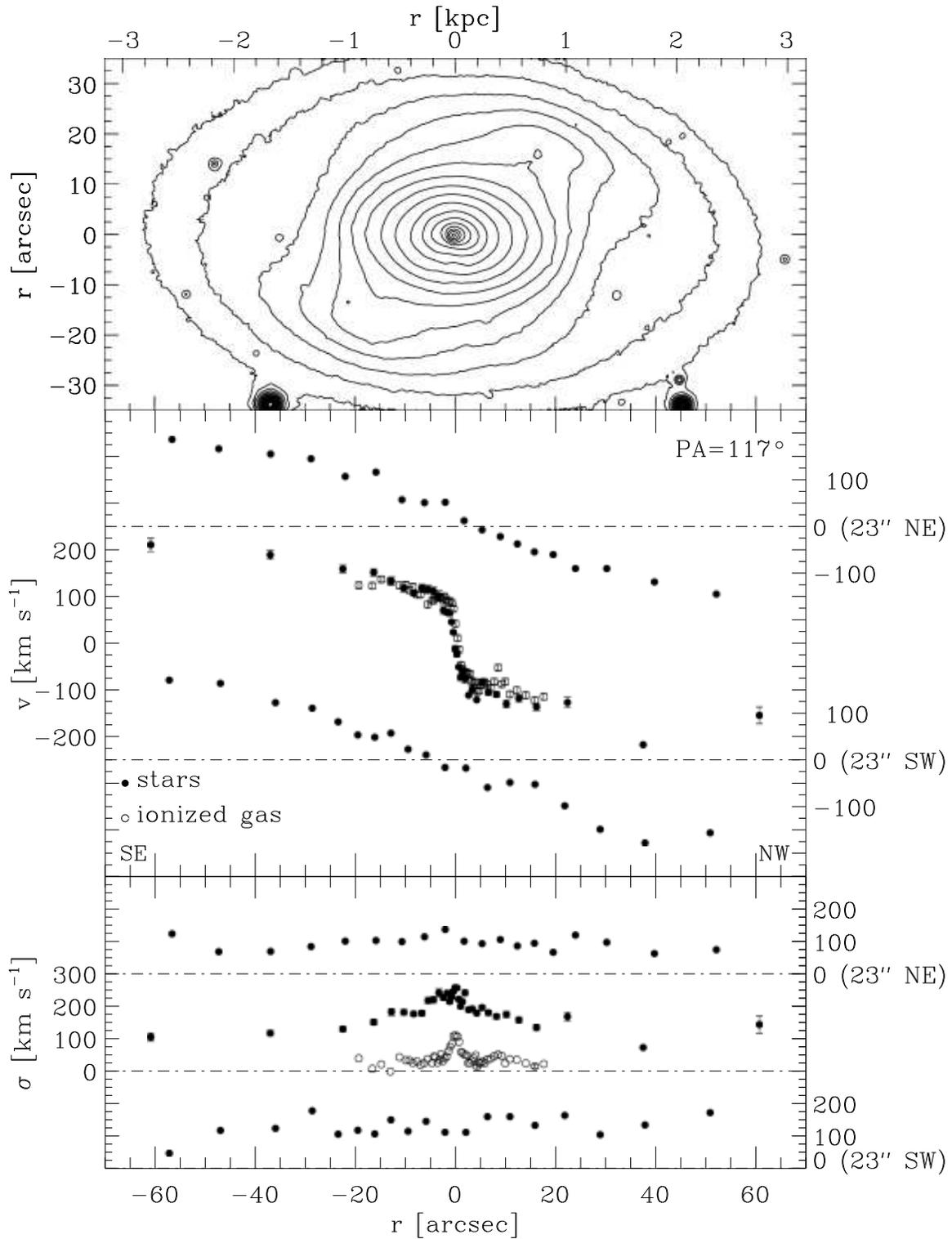}
%\epsscale{0.95}
%\plotone{f2.eps}
\end{center}

\caption{Stellar kinematics for NGC~2787, from three parallel slit
observations at PA =117\arcdeg{} (close to the major axis), taken from
archival WHT-ISIS data.  The isophotes are WIYN $R$-band, rotated to
make PA = 117\arcdeg{} horizontal.
\label{fig:n2787kin}}

\end{figure}

For NGC~2787, we were able to find raw spectroscopic data in the Isaac
Newton Group Archive, from observations made with the ISIS
spectrograph on the William Herschel Telescope in 1995.  These spectra
were taken at three parallel slit positions, all at PA = 117\arcdeg{}
(close to the major axis): one through the nucleus and two offset
spectra (Figure~\ref{fig:n2787kin}).  The reduction and analysis of
these data are discussed below.

\begin{figure}
\begin{center}
\includegraphics{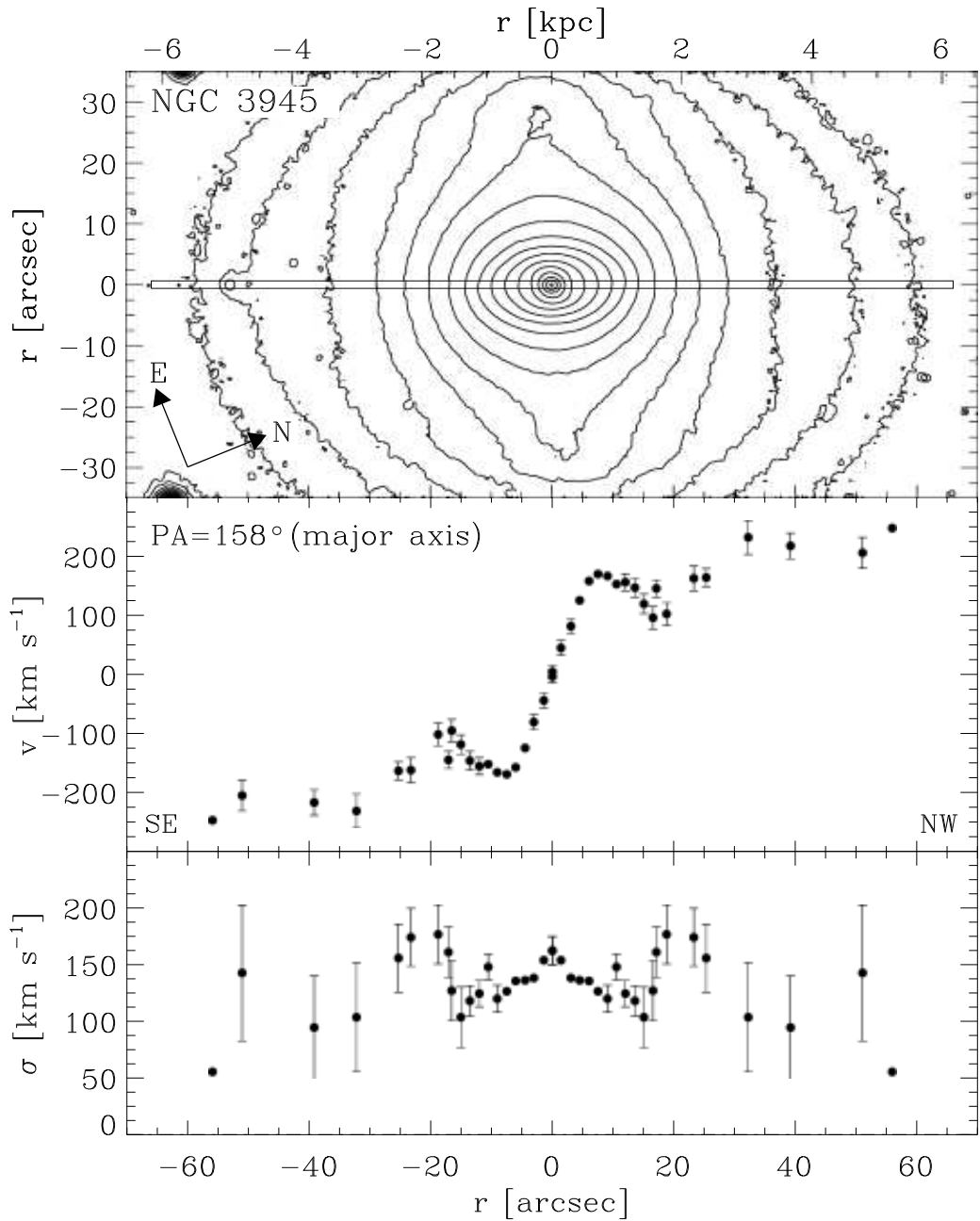}
%%\epsscale{0.9}
%\plotone{f3.eps}
\end{center}

\caption{Stellar kinematics for NGC~3945.  The velocities and velocity
dispersions are taken from the \textit{folded} rotation curve in
\citet{kormendy82a}, so the approaching and receding parts are simply
mirror images; the approaching/receding orientation is based on the
two-sided rotation curve in \citet{bertola95}.  The isophotes are WIYN
$R$-band, rotated to make the major axis horizontal.
\label{fig:n3945kin}}

\end{figure}

For NGC~3945, we relied on the observations of \citet{kormendy82a},
derived from his Figure~3 and Table~5.  The major-axis (PA =
159\arcdeg) velocity and velocity dispersion data points from the
figure were averaged for each radius, with appropriate propagation of
the errors (Figure~\ref{fig:n3945kin}); the spectrum was obtained with
seeing of 1--2\arcsec{}.  A second set of data is the stellar and
gaseous kinematics published by \nocite{bertola95}Bertola et al.\ 1995
(kindly made available to us by Enrico Corsini); however, these go out
to only $r \ap 20\arcsec$ and appear to have much lower S/N than the
Kormendy data.  They \textit{are} useful for determining the kinematic
orientation of the galaxy (i.e., which side is approaching), since the
Kormendy data are folded curves, plotted as functions of radius only.

\subsubsection{Spectroscopic Data Reduction} %[x]

After retrieval from the archives, the NGC~2787 spectra were bias
subtracted, flat-field corrected, cleaned of cosmic rays and
wavelength calibrated using standard MIDAS\footnote{MIDAS is developed
and maintained by the European Southern Observatory.} routines. 
Cosmic rays were identified by comparing the counts in each pixel with
the local mean and standard deviation (as obtained from the Poisson
statistics of the photons, knowing the gain and readout noise of the
detector), and then corrected by interpolating a suitable value.

The instrumental resolution was determined as the mean of the Gaussian
full-width half-maxima (FWHM) measured for a dozen unblended arc-lamp
lines distributed over the whole spectral range of a
wavelength-calibrated comparison spectrum.  The mean FWHM of the
arc-lamp lines, as well as the corresponding instrumental velocity
dispersion, are given in Table~\ref{tab:setup}.  Finally, the
individual exposures of the same slit position were aligned and
coadded using their stellar-continuum centers as reference.  For each
resulting spectrum the center of the galaxy was defined by the center
of a Gaussian fit to the radial profile of the stellar continuum.  The
contribution of the sky was determined from the edges of the resulting
spectrum and then subtracted; NGC~2787 is small enough relative to the
ISIS slit (slit length $\approx 4\arcmin$) that galaxy contamination
is not a problem.

\subsubsection{NGC~2787: Measuring stellar and ionized gas kinematics} 
\label{sec:measuring_kinematics} %[x]

The stellar kinematic parameters were measured from the absorption
lines present on each spectrum using the Fourier Correlation Quotient
Method \citep{bender90}, as applied by \citet{bender94}.  The spectrum
of the K0 star HD 68771, observed during the same night, provided the
best match to the galaxy spectra, and was used as template for
measuring the stellar kinematic parameters of the galaxy.  For each
spectrum we measured the radial profiles of the heliocentric stellar
velocity \vs, velocity dispersion \sigstar, and the Gauss-Hermite
coefficients \htre\ and \hqua\ (the latter two parameters where there
was sufficiently high S/N).  The stellar kinematics are tabulated in
Tables~\ref{tab:ma-kin} and \ref{tab:offset-kin}.
 
The ionized gas kinematics were obtained using the \oiii\ emission. 
The position, FWHM, and uncalibrated flux of the emission line were
individually determined by interactively fitting a single Gaussian to
each emission line and a polynomial to its surrounding continuum,
using the MIDAS package ALICE. The wavelength of the Gaussian peak was
converted to velocity and the standard heliocentric correction was
applied to obtain the ionized gas heliocentric velocity \vg.  The
Gaussian FWHM was corrected for the instrumental FWHM and then
converted to velocity dispersion \siggas.  At some radii where the
intensity of the emission lines was low, we averaged adjacent spectral
rows to improve the signal-to-noise ratio.  The ionized-gas kinematic
parameters of NGC~2787 are listed in Table~\ref{tab:ma-kin}.  We
derived the heliocentric systemic velocity as the velocity of the
center of symmetry of the rotation curve of the gas.

\section{Morphology and Ellipse Fits: Inner Disks as Geometric 
Disks}\label{sec:geometric} %[x]

One of the aims of this study is to check whether the ``inner disk''
identifications tentatively made by
\citet{erwin-sparke02,erwin-sparke03} make sense: are these inner
structures actually disk-like?  Erwin \& Sparke distinguished inner
disks from inner bars by requiring that the position angle differ from
that of the outer disk by less than 10\arcdeg, \textit{and} that the
maximum isophotal ellipticity be less than that of the outer disk. 
This classification is inherently ambiguous, and will sometimes
include inner bars which happen to be aligned with the (projected)
outer disk, because the influence of bulge isophotes can produce a low
apparent ellipticity for an inner bar.  Thus, we would like to be more
certain about the actual geometry of the supposed inner disks of
NGC~2787 and NGC~3945.  In this section, we show that, at least
geometrically, the evidence favors the inner structures being disks
rather than bars: they are most likely flat, approximately circular
structures in the galaxy plane.

\begin{figure}
\begin{center}
%% first is single-page; second is two-column
%\includegraphics[scale=0.95]{ellipse_fits_n2787}
%\epsscale{0.95}
%\plotone{f4.eps}
\includegraphics[scale=0.75]{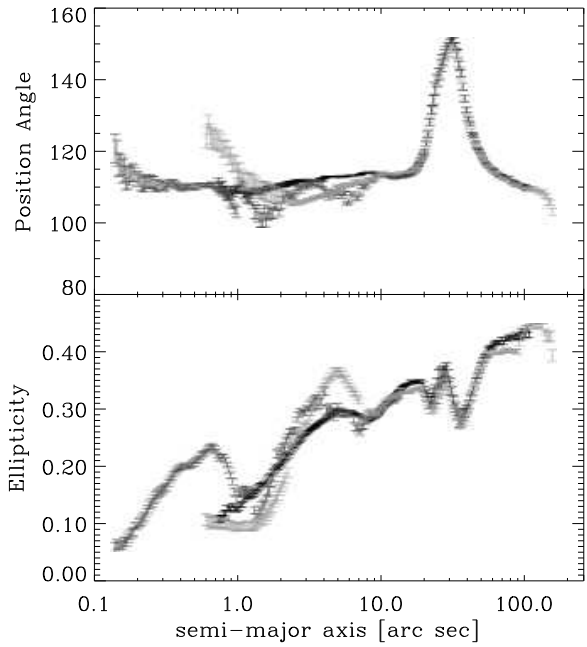}
\end{center}

\caption{Ellipse fits for NGC~2787.  Fits are to WIYN $R$-band (light
gray), \textit{HST} F814W (medium gray), and WHT-INGRID $H$ (black)
images.  The strong variations in the F814W and $R$-band fits (and
weak variations in the $H$-band fits) for $a < 10\arcsec$ are due to
strong dust lanes.}\label{fig:efits-n2787}

\end{figure}

\begin{figure}
\begin{center}
%% first is single-page; second is two-column
%\includegraphics[scale=0.95]{ellipse_fits_n3945}
%\epsscale{0.95}
%\plotone{f5.eps}
\includegraphics[scale=0.75]{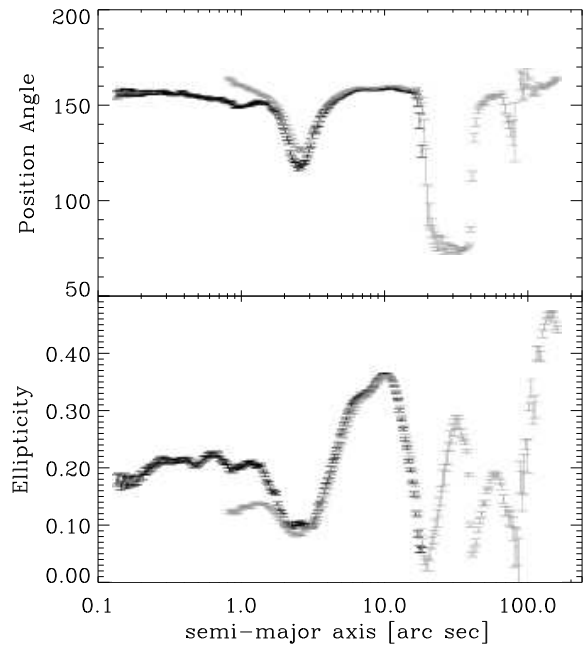}
\end{center}

\caption{Ellipse fits for NGC~3945.  Fits are to WIYN $R$-band (light
gray) and \textit{HST} F814W (black) images.}\label{fig:efits-n3945}

\end{figure}

To measure the orientations and shapes of each galaxy's isophotes, we
used the \textsc{iraf} task \texttt{ellipse}, which is based on the
algorithm of \citet{jz87}.  The position angle and ellipticity of the
fitted ellipses, as a function of their semi-major axis, are shown in
Figures~\ref{fig:efits-n2787} and \ref{fig:efits-n3945}.  Bright stars
and other features have been masked out, including the reflected
starlight which overlies the $R$-band image of NGC~2787 in the outer
disk region.  For outer structures such as the large-scale bars and
outer disks or rings, we use the $R$-band images; for the inner region
where the inner disks dominate, we use the WFPC2 F814W images for
their superior spatial resolution (they are also slightly less
affected by dust extinction).  We also use the INGRID $H$ image of
NGC~2787; although the resolution (Moffat FWHM $\approx 3.5\arcsec$,
measured from bright stars in the same image) is not as good as that
of the \textit{HST} images, it eliminated most of the dust extinction
plaguing the optical images.

For NGC~3945, the maximum ellipticity in the inner zone is $\emax =
0.36$ at a semi-major axis of $a = 10\arcsec$, and the PA of the
ellipse major axis is 158\arcdeg.  The true ellipticity of the inner
disk might be higher, since the isophotes include a contribution from
the bar outside.  \citet{erwin99} found this PA to be essentially
identical to that of most of the other components in the galaxy: the
central bulge (156\arcdeg), the projected lens (156\arcdeg), and the
outer ring (160\arcdeg).  The measured ellipticity implies an
inclination of 52\arcdeg{} for the inner disk, assuming that it has an
intrinsic axis ratio of 0.2 \citep[typical for galactic disks;
e.g.,][]{lambas92}, or 50\arcdeg{} for a geometrically thin disk.  Is
this similar to the overall orientation of the whole galaxy?  The
derivation of NGC~3945's inclination is in fact rather complex, as
there is no visible outer disk.  \citet{erwin99} used arguments based
on measured optical and \hi{} velocities, and on the axis ratios of
the inner ring, lens and outer ring, to argue for an inclination of
around 50\arcdeg, as opposed to 26\arcdeg{} proposed previously by
\citet{w95}; this value is very close to $i = 51\arcdeg$ estimated by
\citet{kormendy82a}.  Thus, the evidence strongly suggests that the
inner feature is geometrically disk-like and coplanar with the galaxy
as a whole.

For NGC~2787, which \textit{does} have a clear outer disk, there is
less scope for ambiguity.  The outer disk's ellipticity, 0.41, implies
an inclination of $\ap 55\arcdeg$ for the galaxy, with a line of nodes
at $\ap 109\arcdeg$.  The ellipticity and position angle of the inner
disk are not easy to measure in the optical images, because of the
presence of strong dust lanes from the off-plane gas.  Using the
$H$-band image, we find an ellipticity of 0.34 and PA $\ap
113\arcdeg$.  This ellipticity suggests an inclination of
49--50\arcdeg, if the inner structure is indeed a disk.  As we show
below, the bulge in this galaxy is relatively large and luminous, so
the measured isophotes are probably rounded due to the bulge
contribution, and the derived inclination for the inner disk is best
seen as a lower limit.  Again, the evidence suggests that what
\citet{erwin-sparke03} called an inner disk \textit{is} probably
disk-like in at least the geometric sense.

Thus, there are good \textit{geometric} reasons for considering these
two features to be disks.  In the following sections, we show that
these inner structures are photometrically and kinematically disk-like
as well.

\section{Photometry, Structure, and Bulge-Disk Decomposition: 
Evidence for Exponential Disks}
\label{sec:decomp}

In Figure~\ref{fig:umasks} we compare unsharp masks of the inner
regions of our two inner-disk galaxies with an unsharp mask of an
inner \textit{bar}.  Bars often produce a clear double-lobed structure
in unsharp masks \citep[see, e.g., Figures~4 and 5 of
][]{erwin-sparke03}, roughly corresponding with the sharp ends of the
bar.  The inner disks, however, produce smooth elliptical features in
the unsharp masks.  Why is their appearance different?  One way to
understand this --- and to see how the inner disks of NGC~2787 and
NGC~3945 are structurally different from typical inner bars --- is to
plot profiles along the major axes of the inner components.  (These
unsharp masks also demonstrate that the ``inner disk'' features in the
ellipse fits are \textit{not} the result of stellar rings, since such
rings are easily visible in unsharp masks; see \nocite{erwin99}Erwin
\& Sparke 1999 and \nocite{erwin01}Erwin, Vega Beltr\'an, \& Beckman
2001).

\begin{figure}
\begin{center}
\includegraphics{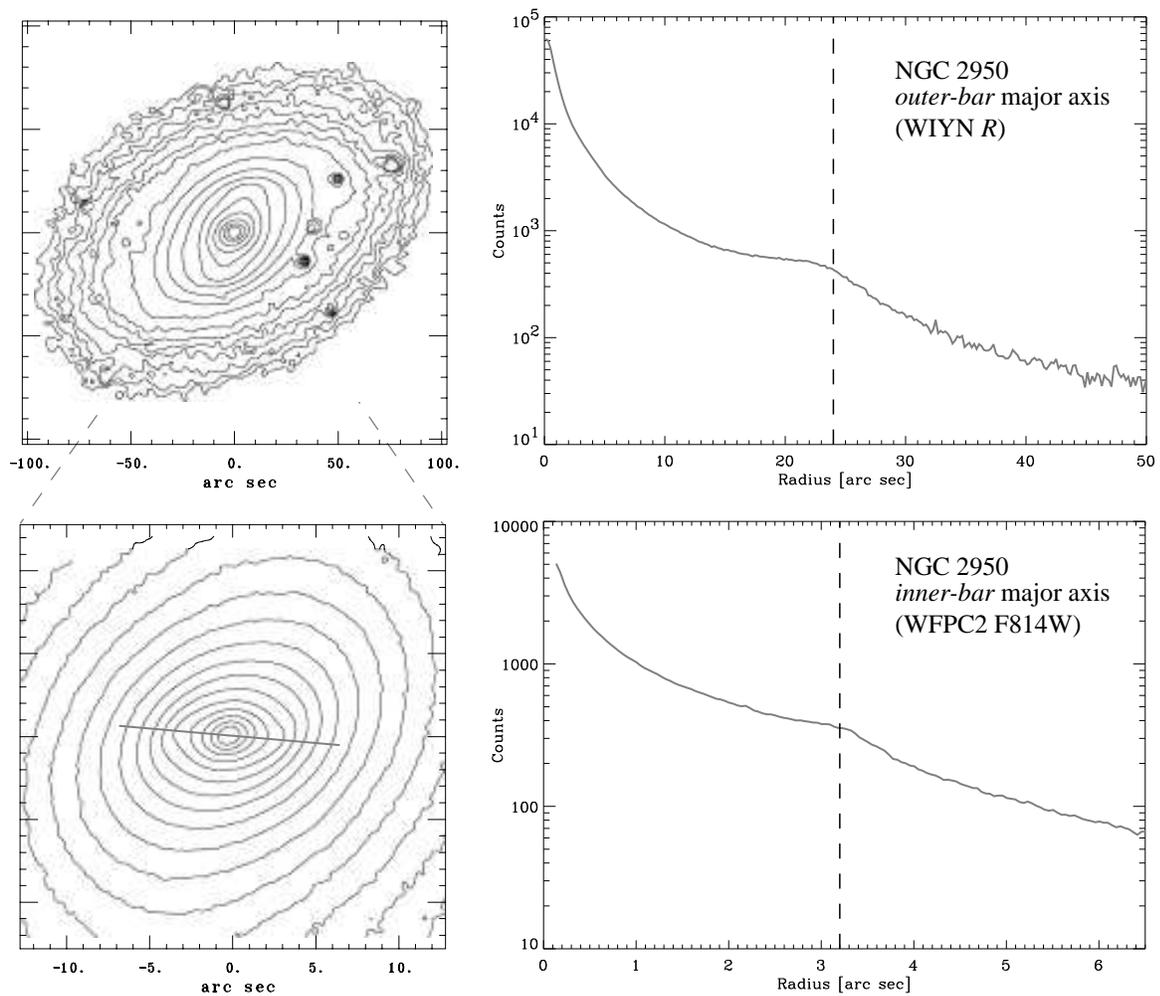}
%\plotone{f6.eps}
\end{center}

\caption{Major-axis profiles along the outer (top) and inner (bottom)
bars of the double-barred galaxy NGC~2950.  The profile for the inner
bar (lower-right panel) is taken from a WFPC2 F814W PC image, to
ensure better resolution; the other figures are all from ground-based
$R$-band images.  All profiles have been folded about the galaxy
center.  The vertical dashed lines indicate the semi-major axis of
maximum ellipticity for each bar.\label{fig:bar-profiles}}

\end{figure}

\begin{figure}
\begin{center}
\includegraphics{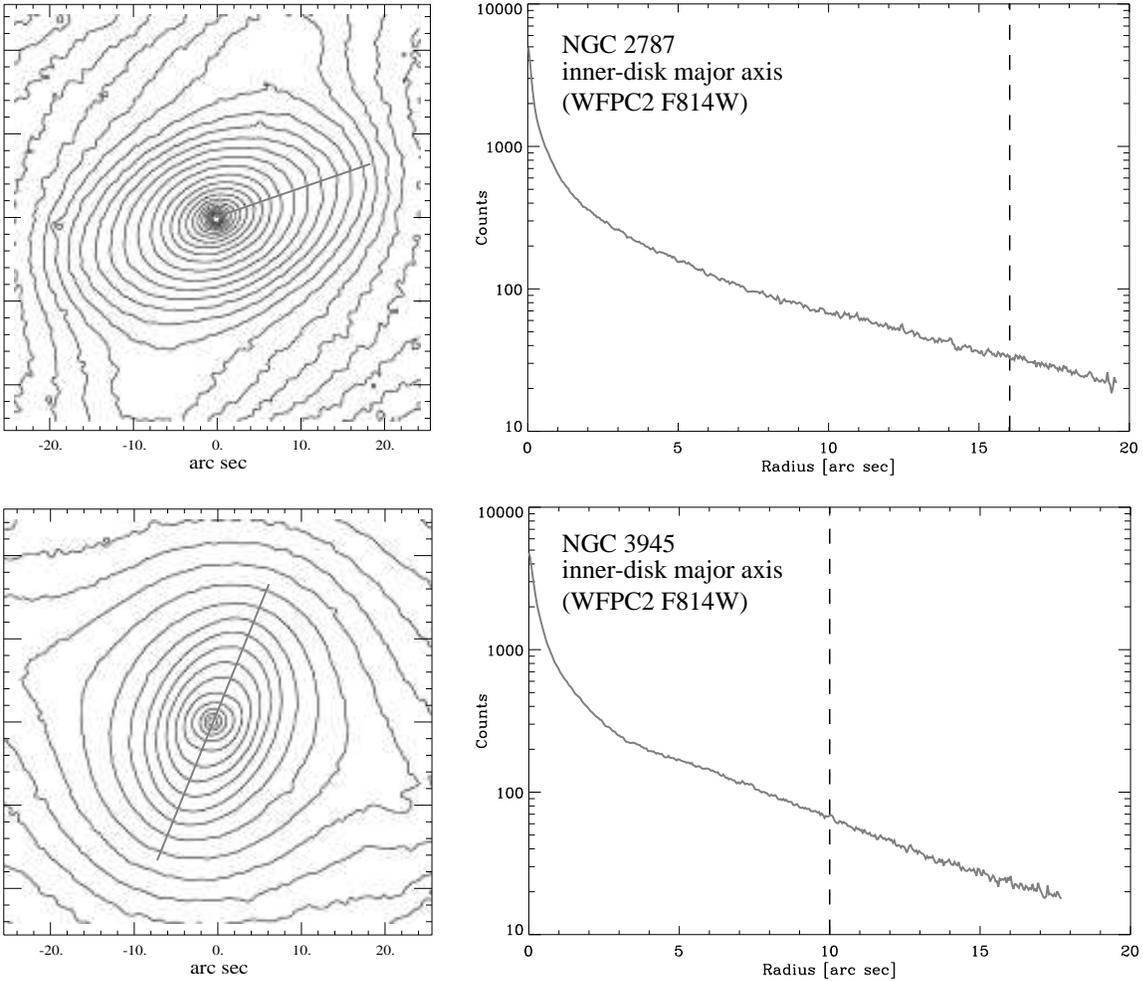}
%\plotone{f7.eps}
\end{center}

\caption{Major-axis profiles along the inner disks of NGC~2787 (top)
and NGC~3945 (bottom).  The isophote plots are $H$-band (NGC~2787) and
$R$-band (NGC~3945); the profiles are taken from the WFPC2 F814W
images, to ensure better resolution.  The profile for NGC~2787 is from
the NW side only, to avoid strong dust lanes on the SE side; the
NGC~3945 profile has been folded about the galaxy center.  The
vertical dashed lines indicate the semi-major axis of maximum
ellipticity for each inner disk.\label{fig:disk-profiles}}

\end{figure}

In Figures~\ref{fig:bar-profiles} and \ref{fig:disk-profiles} we
compare major-axis profiles of the outer and inner bars in the
double-barred galaxy NGC~2950 with major-axis profiles of our two
inner disks.  These profiles are extracted from 3-pixel-wide cuts
along the major axes of each feature.  For NGC~3945 and the bar
profiles of NGC~2950, these profiles are then folded about the galaxy
center.  This last step was \textit{not} done for NGC~2787 because of
the strong extinction due to the off-plane dust lanes.  Fortunately,
the dust lanes are mostly to the SE of the galaxy center; the NW
profile plotted in Figure~\ref{fig:disk-profiles} is relatively free
of extinction.  In each figure, we also indicate the location of
maximum ellipticity from the ellipse fits.

The bar profiles are typical examples of the classic ``flat bar''
shape \citep[e.g.,][]{kormendy82b,ee85}: a shallow profile which breaks
near the radius of maximum ellipticity, with a steep decline outside
into (in the case of primary bars) the outer disk.  From this we can
see that the profiles of the inner disks of NGC~2787 and NGC~3945 do
\textit{not} resemble those of typical bars.

What is striking is how well the inner-disk profiles suggest a classic
bulge-disk structure: a small inner bulge, and an exponential ``disk''
outside.  One way to test this idea is to try decomposing these
profiles into bulge and disk components.  The success of this process
is an indication of how well the bulge + disk hypothesis works; it
also holds the promise of giving accurate measurements for both
components, which will tell us more about the nature of the inner
disks \textit{and} help us understand what kind of bulges these
galaxies really have.  At this stage the decomposition is merely a
hypothesis: we cannot exclude the possibility that these inner disks,
while geometrically disk-like, have complex surface-brightness
profiles.  However, we will argue below that there is some geometric
evidence from the ellipse fits that our decompositions are identifying
two distinct components: the inner disk itself, and a rounder bulge
embedded within it.

We model the profiles as the sum of an exponential profile (for the
disk) and a \citet{sersic68} $r^{1/n}$ profile:
\begin{equation}
  I(r) \, = \, I_{e} \exp \left\{-b_{n} 
  \left[\left(\frac{r}{r_e}\right)^{1/n} - 1 \right] \right\},
\end{equation}
where $I_{e}$ is the surface brightness at the half-light radius
$r_{e}$ and $n$ is the shape parameter; $b_{n}$ is a function of $n$
\citep[see, e.g.,][]{graham01-bulges}.  Each component was
individually convolved with the same Moffat point-spread function
(obtained by fitting stars in the PC images), and the resulting
profiles were added linearly together to form a single profile, which
was then matched to the observed galaxy profile.  A standard
non-linear least-squares algorithm was used until convergence on the
smallest $\chi^2$ value was reached; all data points were weighted
equally.  For NGC~2787, we also added a small point source, for which
there is some evidence from previous studies \citep{sarzi01,peng02};
this improved the fit considerably in the inner arc second of the
profile.  This was not necessary for NGC~3945 (there is some evidence
in the residuals for a nuclear point source, but it appears to be very
weak and does not affect the fits).  The fits were done to the
profiles extracted from the F814W PC images, as described above.

%% Removed footnote:
% \footnote{Weighting by photometry errors tends to
% overemphasize points at small radii, where the S/N is better.}
%
% We also fit a folded profile
% from the F555W image of NGC~3945; we did not do this for NGC~2787
% because the dust lanes near the nucleus are much stronger than in the
% F814W image.

\begin{figure}
\begin{center}
%% first is whole-page version; second is two-column
%\includegraphics[scale=0.9]{ipc_alisterfits}
%\epsscale{0.8}
%\plotone{f8.eps}
\includegraphics[scale=0.5]{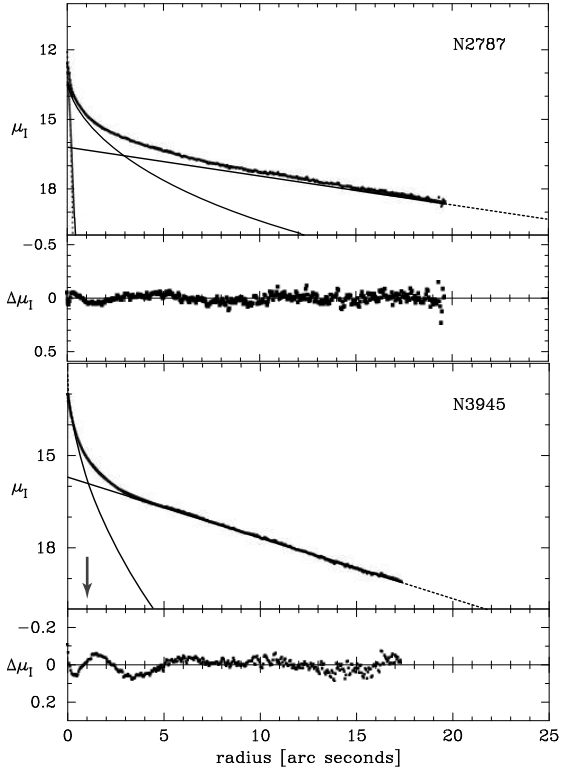}
\end{center}

\caption{Bulge-disk decomposition of the major axis WFCP2 F814W
profiles from Figure~\ref{fig:disk-profiles} of NGC~2787 (top) and
NGC~3945 (bottom), using a seeing-convolved S\'ersic bulge +
exponential disk (+ a central point-source for NGC~2787).  The
surface-brightness profiles have been corrected for Galactic
extinction.  The arrow indicates the outer limits of ``bulgelike''
isophotes in NGC~3945 (i.e., inside this point, the isophotes are
parallel with the outer disk but much rounder; see
Fig.~\ref{fig:efits-n3945}).\label{fig:ipcfits}}

\end{figure}

The resulting fits (Figure~\ref{fig:ipcfits} and
Table~\ref{tab:decomp}) are an excellent confirmation of the
exponential + S\'ersic hypothesis.  For NGC~3945, we measure an
exponential (``disk'') scale length of 5.5\arcsec{} (530 pc).  The
inner S\'ersic region, which we will refer to as the ``bulge,'' is
quite small and is neither exponential nor de Vaucouleurs in profile;
it has S\'ersic index $n = 1.85$ and half-light radius $r_{e} =
1.15\arcsec = 110$ pc.  The decomposition is rather clean; the small
residuals for $r < 5\arcsec$, all less than 0.1 mag, are probably due
to the inner bar and nuclear ring \citep[which affect the isophotes for
$r \approx 1.5$--6\arcsec;][]{erwin99}.

%% 
 % Inner bar sizes, deprojected: 3.9\arcsec, 4.5\arcsec.  This is smaller 
 % than the inner-disk scale length; $\amax/h = 0.7$.  Is this unusually 
 % small compared to typical bar--outer-disk relative sizes?  (I.e., is 
 % the inner-disk--inner-bar system similar to a large-scale galaxy, or 
 % not?)
 %
 % 2787: bulge profile is = disk at 3 arcsec and 2 magnitudes below disk
 % at a = 9 arcsec
 %
 % 3945: bulge profile is = disk at 1 arcsec and 2 magnitudes below disk
 % at a = 3 arcsec
 %%

\begin{figure}
\begin{center}
%% first is whole-page version; second is two-column
%\includegraphics{n2787_disktruncate}
%\plotone{f9.eps}
\includegraphics[scale=0.45]{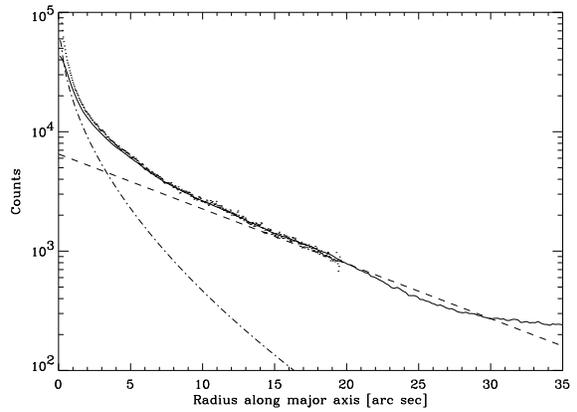}
\end{center}

\caption{$R$-band profile along the major axis of NGC~2787 (solid)
line, along with scaled major-axis profile from WFPC2 F814W image
(dots).  The dashed and dot-dashed lines are the scaled inner-disk and
bulge components, respectively, from the fit to the F814W profile in
Figure~\ref{fig:ipcfits}.  Note that the extrapolated disk profile is
\textit{above} the actual $R$-band light for $r \approx
20$--28\arcsec, indicating that the inner disk is truncated.  (Any
attempt to model the inner profile as a pure bulge would imply a
truncated \textit{bulge} model, which is less plausible than a
truncated disk.)  \label{fig:n2787truncate}}

\end{figure}

For NGC~2787, we find a disk scale length of 8.8\arcsec{} (320 pc). 
The bulge component here is larger than, but structurally similar to,
that of NGC~3945: S\'ersic index $n = 2.31$ and $r_{e} = 4.3\arcsec
\ap 155$ pc.  It is also possible to derive a bulge-dominated fit to
the NGC~2787 profile, with $n \approx 4$ and $r_{e} \approx
40\arcsec$, though the fit is not as good.  There are two major
problems with a bulge-dominated fit for NGC~2787 (beyond the simple
fact that the disk+bulge fit is better).  First, it implies that the
bulge undergoes rapid and significant flattening outwards, which seems
implausible; second, if the fitted bulge model is extrapolated outside
the inner-disk region, it ends up having a surface brightness higher
than that of the galaxy's major-axis profile
(Figure~\ref{fig:n2787truncate}).  (The same is true for the
extrapolated disk in the disk-dominated fit, but disk truncation is
dynamically much more plausible; see Section~\ref{sec:origins}.) 
Thus, to explain NGC~2787's inner disk as purely a bulge, it would
have to flatten rapidly to almost disk-like dimensions \textit{and}
have a sharp radial cutoff, something unknown in other bulges.  The
disk + bulge hypothesis is a simpler and more physically plausible
explanation.

\citet{peng02} fit the inner region of an F547M WFPC2 image of
NGC~2787 using a two-dimensional method.  Their preferred fit was with
a central Gaussian for the nuclear point source and a combination of a
``Nuker-law'' profile \citep{lauer95} with $\epsilon = 0.23$ and an
exponential with $\epsilon = 0.40$ and scale length = 7.1\arcsec. 
However, they noted that ``the bulge component can also be well fitted
by a superposition of a S\'ersic ($m_{\rm F547M} = 12.43$ mag, $n =
2.71$, and $r_{e} = 7.43\arcsec$) and an exponential ($m_{\rm F547M} =
12.43$ mag and $r_{s} = 11.7\arcsec$).''  Their two exponential scale
length measurements bracket ours, and their S\'ersic fit is probably
consistent with ours \citep[typical errors for $n$ in S\'ersic fits
are $\sim 25$\%;][]{caon93}.\footnote{The ``luminous stellar disk at
the center of the dust disk'' which Peng et al.\ find in their
residuals is almost certainly, given its size and orientation, a side
effect of the sharp cutoff in dust extinction at $a \ap 1.5\arcsec$,
rather than a distinct \textit{stellar} feature.}

%% 
 % Also give 2787 outer-disk exponential fit, just 'cause we have it (and
 % compare with BBA?)
 % 
 % 
 % PERHAPS: What is intrinsic shape of NGC~3945 disk?  One possible
 % angle: find scale length (from B-D decomposition, or just from fits to
 % outer region) for several position angles: major axis, and also +/- 5
 % or 10 degrees.  Small deviations from major axis should minimize bar
 % contamination, hopefully; using change in $h$ with angle should allow
 % us to define its observed ellipticity better, and thus determine
 % inclination of disk (assuming it is circular), or at least possible
 % intrinsic shapes assuming various inclinations.
 %%

Our analysis so far seems to suggest that the inner light of these
galaxies can be decomposed into two components: a S\'ersic bulge and
an exponential disk.  Such a photometric decomposition of a
one-dimensional profile does not, however, guarantee that there are
two distinct morphological components: while stellar disks are often
exponential, there is no reason they must be.  Could we be dealing
with disks whose surface-brightness profiles change with radius, but
which remain flat and disk-like into the limits of resolution?  In
fact, there is morphological evidence that this is not the case, and
that there \textit{are} distinct, albeit small, bulges in both
systems.

The ellipse fits to the F814W image of NGC~3945
(Figure~\ref{fig:efits-n3945}) clearly show that the isophotes of the
inner $r \lesssim 1\arcsec$ are parallel with the inner disk and the
outer ring, but much rounder.  This is consistent with the innermost
light being dominated by a mildly oblate component like a bulge.  And
this is precisely the region dominated by the S\'ersic component in
our fit (Figure~\ref{fig:ipcfits}).  Thus, it is plausible to argue
that the inner light of NGC~3945 comes from both a bright inner disk
with an exponential profile (with, as well, a nuclear ring and inner
bar) and a compact, approximately oblate non-exponential bulge.

The case for NGC~2787 is more complicated, due to the effect of the
dust lanes.  Nonetheless, the innermost isophotes in the F814W image
--- at $r < 1\arcsec$, interior to the dust rings --- have
ellipticities $\lesssim 0.2$ (Figure~\ref{fig:efits-n2787}), which is
not consistent with a pure disk morphology.  Unfortunately, the INGRID
$H$-band image --- where the dust rings have the least influence ---
is too low in resolution to be useful here: the ellipse fits indicate
that the isophotes get steadily rounder towards the center, but this
is at least partly due to seeing effects.

How far out do the exponential disks extend?  For NGC~2787, there is
evidence for a truncation at $r \sim18$--20\arcsec --- about 2.5 scale
lengths --- as can be seen in Figure~\ref{fig:n2787truncate}.  There
is a steeper falloff outside this point, leading to the much shallower
profile of the lens at $r \sim 27$--30\arcsec.  While it might be
argued that this falloff is a signature of the end of a \textit{bar},
thus suggesting that the inner disk is an inner bar after all, this is
unlikely: the maximum ellipticity (at $a = 16\arcsec$ in $K_{s}$) is
well \textit{inside} the turnover point, which is not typical of bars. 
Moreover, if this turnover does mark the end of an inner bar, then
this inner bar is $\approx 50$\% the size of the outer bar, which is
twice the relative size of the largest confirmed inner bars
\citep{erwin03-db}.  For NGC~3945, there is no clear truncation; the
profile merges with that of the lens at $r \ap 17\arcsec$.  This is
also the point in the ellipse fits where the position angle starts
twisting away towards that of the outer bar, which suggests that the
disk has, if not actually ceased to exist, become too faint enough to
affect the isophotes.  So the inner disk in this galaxy can be traced
to about 3 scale lengths; it may or may not be truncated.  There are
dynamical reasons to expect that the disks should be truncated at some
point, if the disks are supported by $x_{2}$ orbits of the bars (the
outer bar in the case of NGC~3945), since these orbits do not extend
outside the bar region; see Section~\ref{sec:origins}.

\begin{figure}
\begin{center}
%% first is whole-page version; second is two-column
%\includegraphics[scale=0.9]{dejong_disks}
%\epsscale{0.9}
%\plotone{f10.eps}
\includegraphics[scale=0.45]{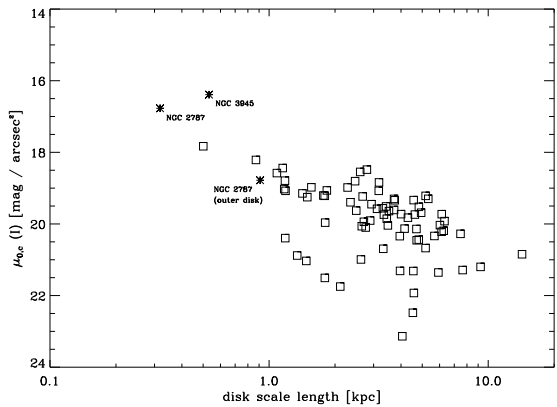}
\end{center}

\caption{Disk central surface brightness versus scale length in $I$. 
The squares are outer-disk values, from the S\'ersic + exponential
fits by Graham (2001a; see also Graham 2001b) to the azimuthally
averaged profiles of \citet{dejong94}.  The inner disks studied in
this paper are indicated by stars, along with the outer disk of
NGC~2787 (NGC~3945 does not have an exponential outer disk).  All
central surface brightnesses have been corrected for Galactic
extinction, inclination, and transparency, as described in the text.
\label{fig:dejong-disks}}

\end{figure}

These inner disks are smaller and have higher central surface
brightnesses than is typical for large galactic disks. 
Figure~\ref{fig:dejong-disks} compares our two inner disks with the
$I$-band scale lengths of (outer) disks from the sample of
\citet{dejong94}.  The disk scale lengths are from the
seeing-corrected, S\'ersic + exponential fits of
\citet{graham01-bulges} to the azimuthally averaged profiles of de
Jong \& van der Kruit, corrected for inclination using the axis ratios
from \nocite{rc3}de Vaucouleurs et al.\ (1991, hereafter RC3) and for
transparency as described in Section~\ref{sec:luminosity}.  The
physical scale lengths use kinematic distances from LEDA, as described
in Section~\ref{sec:overview}.  (We also include approximate disk
parameters for the \textit{outer} disk of NGC~2787, based on an
exponential fit to major axis profile, outside the bar region, from
the F814W mosaic image.)  It can be seen that the inner disks of
NGC~2787 and NGC~3945 are outside the range of most of the large
disks, though they appear to form a continuation of the $\mu_{0}$--$h$
trend or upper envelope (smaller disks tend to have higher central
surface brightnesses).  In this sense, they are probably part of the
overall $\mu_{0}$--$h$ trend or envelope suggested by, e.g.,
\citet{scorza98} in the $V$ band for a wide range of disks, including
the nuclear disks of ellipticals.

\section{Kinematic Evidence for Disks and Resonances}
\label{sec:kinematics} %[x]

Having established that the inner structures in these two galaxies are
probably disk-like in geometry, and similar to typical large-scale
disks photometrically, we now turn to the issue of whether they are
\textit{kinematically} disk-like.  Lacking detailed, two-dimensional
spectroscopy, we restrict ourselves to asking the following questions:
Do the stellar kinematics appear to be dominated by rotation, as is
typical for disks?  Or are they mixtures of random motions and
rotation consistent with models of rotationally flattened, isotropic
spheroids --- the standard models for bulges?  We also use rotation
curves to look for resonances: since both of the inner disks are found
within strong bars, it is crucial to understand how the disks interact
with bar resonances in order to understand their dynamical nature, and
possibly their origins as well.

\subsection{Comparing Inner Disks with Bulge Models} %[x]

\citet{kormendy82a} first pointed out the kinematic peculiarity of
NGC~3945's ``bulge'' (i.e., what we consider to be the bulge + inner
disk).  In a study of nine barred galaxy bulges, he measured stellar
rotation velocities and velocity dispersions, and then compared these
with predictions from bulge models.  One measure of the relative
importance of rotation versus random motions is the ratio \vms, with
$V_{\rm max}$ being the maximum rotational velocity within the bulge,
and $\sigma$ the central or mean velocity dispersion.  In simple
models for kinematically hot, spheroidal systems --- isotropic oblate
rotators --- there is a direct relationship between \vms{} and the
ellipticity of the bulge \citep{binney78}; this can be approximated as
$\vms \ap \sqrt{\epsilon / (1 - \epsilon)}$ \citep{kormendy82b}.  The
bulges of \textit{unbarred} galaxies studied by \citet{kormendy-ill83}
followed this relation, as did some of the bulges of barred galaxies. 
However, several barred-galaxy bulges --- among them NGC~3945 --- were
notable for having $\vmsstar > 1$, indicating either significant
velocity anisotropies or dominance by rotation.\footnote{\vmsstar{} is
the ratio of the observed \vms{} to the predicted value.}

To calculate \vmsstar{} for NGC~2787 and NGC~3945, we use the
\textit{maximum} measured ellipticities of the inner disks --- 0.35
for NGC~2787 and 0.36 for NGC~3945 --- so the predictions should be
seen as upper limits for the bulge models.  $\sigma$ is taken to be
the mean velocity dispersion within the inner-disk region.  In both
cases, the observed \vms{} values are too high to be explained by the
isotropic oblate rotator model.  For NGC~2787, $\vmsstar \sim 1.4$;
for NGC~3945, the ratio is $\sim 1.6$ (our ratio is lower than
Kormendy's $\vmsstar = 1.8$ because we measure a higher ellipticity
for the inner-disk region, which in turn leads to a higher predicted
\vms{}).

An additional argument against the isotropic oblate rotator model for
NGC~3945 was made by \citet{kormendy82a}, comparing the radial and
azimuthal velocity dispersions (his Table~5 and Figure~4, which use
velocity dispersions from an unpublished minor-axis spectrum).  In
contrast to the barred S0's NGC~936 and NGC~4596, where the ratio
$\sigma_{r}/\sigma_{\phi}$ stays close to 1.0 out to the edge of the
bulge-dominated regions, in NGC~3945 this ratio climbs to almost 1.4
in the inner-disk regions.  As Kormendy pointed out, the true ratio is
probably higher, since $\sigma$ observations along both major and
minor axes are diluted by contributions from the vertical velocity
dispersion $\sigma_{z}$.  Kormendy interpreted this as evidence for
bar kinematics in NGC~3945; however, as we have shown above, the inner
disk is almost certainly \textit{not} barlike.  Moreover,
\nocite{moiseev02}Moiseev's (2002) 2D stellar velocity field of the
inner $7\arcsec \times 7\arcsec$ of this galaxy appear regular and
typical of circular motion.  A $\sigma_{r}/\sigma_{\phi}$ ratio
$\gtrsim 1.4$ \textit{is} typical for G--M giants in the Milky Way
disk \citep[e.g., Table~10.3 in][]{bm98}, again suggesting that we are
seeing disk rather than bulge kinematics in the inner disk of
NGC~3945.

\subsection{Circular Velocities and the Location of Resonances} %[x]

In this section we obtain approximations to the resonance curves of
the two galaxies.  Resonance curves are derived from velocity curves
by calculating the radial variation of angular velocity $\Omega =
V/r$, and then the radial variation of $\Omega - \kappa/2$, where
$\kappa$ is the epicyclic frequency, determined as
\begin{equation}
  \kappa^{2} = r \frac{d \Omega^{2}}{d r} + 4 \Omega^{2}.
\end{equation}
There are several inherent difficulties in relating these idealized
resonance curves to actual resonances in a galaxy, some of which we
discuss below.  The most pressing problem from the observational
standpoint is what type of velocity curve to start with.

In general, gas velocity curves reflect the true circular velocities
much better than stellar velocities.  Unfortunately, we do not have
gas velocities for NGC~3945 (the gas kinematics presented by
\nocite{bertola95}Bertola et al.\ 1995 are noisy and do not extend
beyond the inner $\sim 3\arcsec$; \nocite{moiseev02}Moiseev 2002
suggests that the gas in this region actually counter-rotates with
respect to the stars).  The major-axis spectrum of NGC~2787 shows what
appears to be a good gas rotation curve.  However, additional
emission-line spectra at several different position angles show that
the gas is kinematically misaligned with respect to the stars, with a
line-of-nodes close to 70\arcdeg, in contrast to the optical (stellar)
major axis at 109\arcdeg.  This, combined with the distinctive dust
lanes, indicates that the ionized gas is orbiting in a plane tilted
with respect to the stars, and is in fact counter-rotating; the
gaseous kinematics will be presented and discussed further in
\citet{vega-beltran03}.  Consequently, we must use the stellar
kinematics to estimate the resonance locations.

Even in an axisymmetric disk, the observed major-axis stellar rotation
velocities $V_{\phi}$ (corrected for projection effects) will be lower
than the circular velocities $V_{c}$ because velocity dispersion
provides some dynamical support against gravity.  To recover the
circular velocities, we need to make an asymmetric drift correction. 
Following, e.g., \citet{kormendy84-1553}, the correction in the disk
plane can be written as
\begin{equation}
  V_{c}^{2} - V_{\phi}^{2} = - \sigma_{r}^{2} \left[ \frac{\partial
  \ln \rho}{\partial \ln r} + \frac{\partial \ln
  \sigma_{r}^{2}}{\partial \ln r} + \left(1 -
  \frac{\sigma_{\phi}^{2}}{\sigma_{r}^{2}}\right) \right]
\end{equation}
where $\rho$ is the volume mass density.  This equation
presupposes an axisymmetric, disk-dominated system; it also assumes
that the velocity ellipsoid points towards the galaxy center
everywhere.  Again following Kormendy, we assume that $\rho$ can be
approximated by the surface brightness (this assumption relies on the
disk having an unchanging M/L ratio and a constant scale height; the
latter assumption will obviously start to fail where the bulge
dominates the surface brightness).  For the surface brightness
profiles, we use azimuthally averaged surface-brightness profiles from
the WIYN $R$-band images (which provide the most complete coverage).

For both galaxies, we have $\sigma_{\rm maj}$, the observed velocity
dispersions along the major axis; this is a combination of
$\sigma_{\phi}$ and $\sigma_{z}$.  Ideally, we could determine all
three velocity dispersion components with three position angles, but
outside of a few isolated points, we do not even have measurements of
$\sigma_{\rm min}$.  Unfortunately, it is not trivial to isolate
$\sigma_{\phi}$ from $\sigma_{\rm maj}$ alone, or to determine
$\sigma_{r}$.  Various assumptions about the relation between
components of the velocity dispersion (i.e., the shape of the velocity
ellipsoid) can be made, but these rely on assumptions about the mass
distribution or the circular velocities which are probably not true
over the regions we are interested in.  For example, if the rotation
curve is flat, then the epicyclic approximation predicts
$\sigma_{r}^{2} = 2 \sigma_{\phi}^{2}$.  Observations suggest that
$\sigma_{z} \sim 0.7 \sigma_{r}$, both for the Milky Way
\citep[e.g.,][and references therein]{bm98} and for NGC~488
\citep{gerssen97}; but $\sigma_{r}/\sigma_{\phi}$ varies from 1.25
\citep[NGC~488][]{bm98} to almost 2 \citep[Milky Way dwarfs;][]{bm98}. 
We adopt, as a compromise, $\sigma_{z} = \sigma_{\phi} = 1/\sqrt{2}
\sigma_{r}$, but note that this is probably not a very good
assumption.

To carry out the correction, we fit the observed velocity and velocity
dispersion profiles with smooth interpolation functions.  When the
derivative of the surface-brightness profile is used to calculate the
corrected velocities, it introduces a fair amount of noise. 
Consequently, we also fit the corrected velocities with smooth
interpolation functions in order to calculate the frequencies.

%% 
 % In Figures~xxx\ref{fig:n3945-freq}, we present the results of these
 % calculations.  Figure~xxx shows the (deprojected) rotation velocities,
 % the drift-corrected velocities, and the smooth fits used for
 % calculating derivatives.  The results for NGC~2787 are not
 % unreasonable, and in fact the outer velocities ($\sim 350$ \kms) are
 % only slightly higher than those derived by \citet{neistein99} for the
 % same galaxy (see their Fig.~1).  The corrected velocities at $r
 % \lesssim 5\arcsec$, on the other hand, are probably not reliable,
 % because this is the bulge-dominated region: here, the disk
 % approximations used to derive the drift-corrected velocities are no
 % longer valid, and the observed velocity dispersion is probably a
 % composite of disk and bulge components.
 %%

\begin{figure}
\begin{center}
%% first is whole-page version; second is two-column
%\includegraphics[scale=0.7]{drift_corrections}
%\epsscale{0.7}
%\plotone{f11.eps}
\includegraphics[scale=0.45]{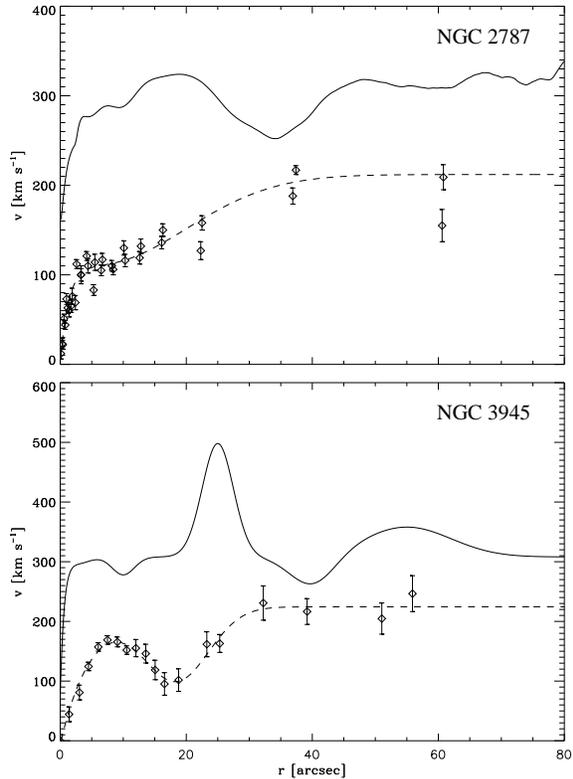}
\end{center}

\caption{Deprojected stellar rotation curves and drift-corrected
curves for NGC~2787 (top) and NGC~3945 (bottom).  The data points are
the deprojected, observed stellar velocities; the smooth curves are
fits to the observed velocities (dashed lines) and fits to the
velocities after correction for asymmetric drift (solid
lines).\label{fig:drift}}

\end{figure}

\begin{figure}
\begin{center}
%% first is whole-page version; second is two-column
%\includegraphics[scale=0.7]{n2787_freq}
%\epsscale{0.7}
%\plotone{f12.eps}
\includegraphics[scale=0.45]{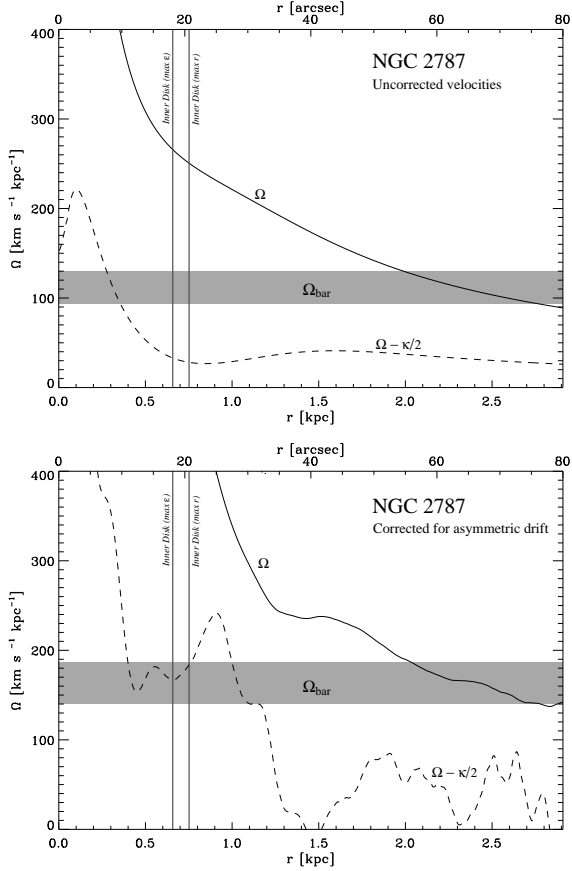}
\end{center}

\caption{Resonance curves for NGC~2787, based on the uncorrected
velocities (top) and the drift-corrected velocities (bottom).  The
$\Omega$ (solid) and $\Omega - \kappa/2$ (dashed) curves are plotted,
along with the estimated bar pattern speed (horizontal gray band); the
intersection of the latter with the $\Omega - \kappa/2$ curve
indicates a possible inner Lindblad resonance.  The vertical dark gray
lines indicate the radii of the inner disk's maximum isophotal
ellipticity and approximate truncation.\label{fig:n2787-freq}}

\end{figure}

\begin{figure}
\begin{center}
%% first is whole-page version; second is two-column
%\includegraphics[scale=0.7]{n3945_freq}
%\epsscale{0.7}
%\plotone{f13.eps}
\includegraphics[scale=0.45]{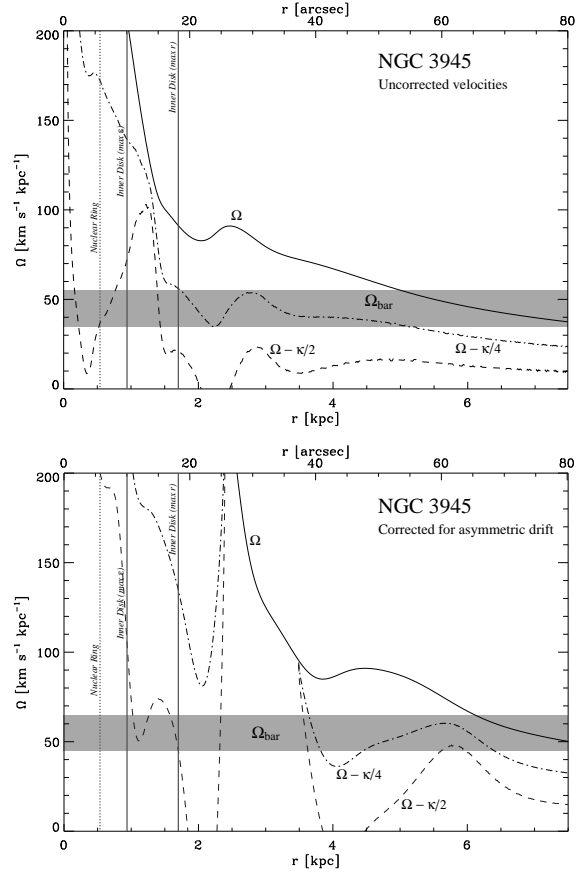}
\end{center}

\caption{As for Figure~\ref{fig:n2787-freq}, but showing the resonance
curves for NGC~3945.  We include the $\Omega - \kappa/4$ curve
(dot-dashed line), since we use it to help estimate the bar pattern
speed (see text).  We also mark the approximate location of the
stellar nuclear ring (vertical dashed line).  In the resonance curve
made from the drift-corrected velocities (lower), curves in the region
$r = 25$--40\arcsec{} are unreliable due to peculiar kinematics in the
lens (see text).\label{fig:n3945-freq}}

\end{figure}

Figures~\ref{fig:drift}--\ref{fig:n3945-freq}, show the
results of these calculations.  Figure~\ref{fig:drift} shows the
(deprojected) rotation velocities, the drift-corrected velocities, and
the smooth fits used for calculating derivatives.  The results for
NGC~2787 are reasonable, and in fact the outer velocities ($\sim
350$ \kms) are only slightly higher than those derived by
\citet{neistein99} for the same galaxy (see their Fig.~1).  The
corrected velocities at $r \lesssim 5\arcsec$, on the other hand, are
probably not reliable, because this is the bulge-dominated region:
here, the disk approximations used to derive the drift-corrected
velocities are no longer valid, and the observed velocity dispersion
is probably a composite of disk and bulge components.

The results for NGC~3945 are somewhat dubious.  The most questionable
part of the drift-corrected curve is the lens region (roughly, $r =
20$--40\arcsec), where the corrected velocity rises to $\sim 500$
\kms.  This is because the velocity \textit{dispersion} rises rather
dramatically within this region (Figure~\ref{fig:n3945kin}), forcing a
large correction.  A similar rise in velocity dispersion can be seen
in the inner part of the unbarred S0 NGC~1553's lens
\citep{kormendy84-1553} (our unpublished major-axis spectrum of the
SB0 NGC~2950 shows a similar rise in $\sigma$ in its lens region), and
a corresponding peak in its drift-corrected velocity is suggested by
Kormendy's Figure~7.  The lens regions are where the the bar dominates
the light; they are thus the part of the galaxy furthest from
axisymmetry, so asymmetric drift corrections --- which assume
axisymmetry --- are probably least reliable here.  As with NGC~2787,
the corrections at small radii ($r \lesssim 3\arcsec$) may not be very
reliable either: though the bulge in NGC~3945 is quite small, the
spatial resolution of the original spectrum was $\sim 2\arcsec$. 
Finally, to specify the curve at $r \gtrsim 60\arcsec$, beyond the
last measured velocity, we assumed a flat rotation curve beyond the
last measured point ($r = 56\arcsec$); since the \hi{} half-width is
$\approx 340$ \kms \citep[$\approx 440$ \kms deprojected;][]{wk86},
the true velocity curve probably continues rising.

Because of the uncertainties in the asymmetric drift correction, we
show in Figures~\ref{fig:n2787-freq} and \ref{fig:n3945-freq}
resonance curves derived from the observed velocities, as well as from
the drift-corrected velocities.  The truth, we hope, may lie somewhere
in between.  Superimposed on these curves are crude estimates of the
bar pattern speed (outer-bar pattern speed in the case of NGC~3945);
the points where this crosses the $\Omega - \kappa/2$ curves indicate
approximate locations of inner Lindblad resonances.  For NGC~2787
(Figure~\ref{fig:n2787-freq}), we estimate the pattern speed by
assuming that the bar ends near corotation (CR), so that CR = 1 to 1.4
$R_{\rm bar}$; this is the range observed for bars in S0 galaxies
where the pattern speed has actually been measured \citep{aguerri03}. 
For $R_{\rm bar}$ we use the upper limit on bar semi-major axis from
\citet{erwin-sparke03}, which is 36\arcsec{} (deprojected $a =
54\arcsec$); we use the drift-corrected resonance curve to estimate
$\Omega_{\rm bar}$.  The result is $\Omega_{\rm bar} \sim 90$--140 or
140--180 \kmskpc, depending on which $\Omega$ curve we use.  These
values are quite high (the pattern speeds found by Aguerri et al.\
range from 40--100 \kmskpc), but might be plausible given both the
high velocities in NGC~2787 and the small size of the bar ($a < 2$
kpc).

For NGC~3945 (Figure~\ref{fig:n3945-freq}), we make the same
assumptions, using $a = 39\arcsec$ (deprojected $a = 61\arcsec$) from
\citet{erwin-sparke03}, though the rotation curves do not go out far
enough for us to set a reliable lower limit on the pattern speed; the
estimate is then $\Omega_{\rm bar} \sim 45$--55 \kmskpc.  We can also
estimate the pattern speed by assuming that the dusty inner ring (at
the rim of the lens, with $a \sim 36\arcsec$ measured along the
primary bar major axis) is near the inner ultra-harmonic resonance
($4:1$ resonance) of the primary bar, where $\Omega - \kappa/4 =
\Omega_{p}$, since simulations and observations both suggest this
connection \citep[e.g.,][]{buta96,buta01}; this yields $\Omega_{\rm
bar} \approx 35$--45 \kmskpc.  Very roughly, then, we can argue that
$\Omega_{\rm bar} \approx 35$--55 \kmskpc{} for NGC~3945.

It is important to remember that resonances derived from rotation
curves are exact only for axisymmetric systems with infinitesimal bar
perturbations.  There are two reasons for this: first, the observed
rotation curves themselves may contain non-circular components induced
by the bar, making it harder to recover the nominal circular-velocity
curve; second, the location of resonances in a strongly
non-axisymmetric potential will usually not be identical with those
derived by assuming an almost axisymmetric potential.  Accurate
identification of resonances in strongly barred galaxies requires a
more complicated approach --- e.g., deriving an accurate potential
from the photometry and using orbit integrations to find the different
resonance-related families, something beyond the scope of this paper. 
However, using a simple approach to the resonance location does give
us a rough guide to help see their approximate positions, and can give
us a prima facie answer to the question of whether the presence of
inner stellar disks is related to, e.g., the presence of inner
Lindblad resonances.  In addition, the presence of the inner disks
makes the the galaxies more axisymmetric in just the regions we are
most interested in, which should reduce the uncertainties.

We \textit{do} find good evidence that \textit{both} galaxies have at
least one ILR, and that this ILR is in the inner disk region.  In the
case of NGC~2787 (Figure~\ref{fig:n2787-freq}), the ILR is at $r \sim
10$--30\arcsec{} and is probably the only ILR, since the $\Omega -
\kappa/2$ curve continues rising inward (the abrupt downturn at $r <
1\arcsec$ in the upper curve is due to seeing effects, and the
existence of a central supermassive black hole \citep{sarzi01}
probably ensures that there is no second ILR at small radii).  For
NGC~3945 the situation is ambiguous: there is very good evidence for
an ILR at $r \approx 15\arcsec$, and there is a suggestion of a second
ILR at $r \sim 5\arcsec$, at least in the resonance curve derived from
the observed velocities.  However, at this point we are getting close
to the spatial resolution of the original spectrum, and there are
probably non-circular motions due to the inner bar, so the resonance
curve in that region is unreliable.

\section{Discussion}

This paper has, in a sense, been directed at examining the argument
made by \citet{kormendy93} in a review article: that at least some
photometric ``bulges'' --- i.e., central light concentrations --- in
early type galaxies are not really classical bulges --- i.e.,
spheroidal, kinematically hot systems --- but are instead part of the
disk.  We chose the galaxies NGC~2787 and NGC~3945 because we had
available the right observations, and our previous studies
\citep{erwin99,erwin-sparke02,erwin-sparke03} had in fact hinted at
this possibility.  Moreover, NGC~3945 was one of the most dramatic
cases of a kinematically disk-like ``bulge'' in the study of
\citet{kormendy82a}.  The most notable result of the present exercise
is the demonstration that most of the inner light in these two
galaxies --- previously ascribed to their bulges --- comes from
luminous, exponential disks embedded \textit{inside} large bars.  This
does \textit{not} mean that these galaxies lack hot, spheroidal bulges
entirely: we do find evidence for such bulges, inside the inner disks. 
But they are smaller (in the case of NGC~3945, \textit{much} smaller)
than a na\"{\i}ve bulge-disk decomposition of the whole galaxy would
suggest, as we show below.

\subsection{Bulge and Inner Disk Luminosities and Masses, and the
Question of Hubble Type}
\label{sec:luminosity}

Since the central light in these galaxies is divided between the bulge
and the inner disks, it is of interest to determine the relative and
absolute luminosities of the inner components.  How luminous (and
massive) are these inner disks?  How much of what might na\"{\i}vely
be considered the bulge is really bulge light, and how much comes from
the inner disk?  What is the true bulge/total luminosity ratio for
these S0 galaxies?  Here, we are explicitly defining \textit{bulge} to
mean a spheroidal or triaxial, kinematically hot stellar component,
often (but not necessarily) with a non-exponential light profile.  As
we point out below, this is emphatically \textit{not} always the same
thing as ``inner luminosity excess above the outer disk profile,'' a 
common definition of bulge which is often assumed to be equivalent.

To compute the luminosities of the inner disks and bulges of these
galaxies, we assume that they are well modeled by the fitted
exponential and S\'ersic functions, respectively.  We assume the inner
disks are truncated at outer radii of 20\arcsec{} and 17\arcsec{} for
NGC~2787 and NGC~3945, respectively (see Section~\ref{sec:decomp}). 
For comparison with the total luminosities of each galaxy, we use the
absolute magnitudes from Table~\ref{tab:general}.

The luminosity of an ellipsoid of ellipticity $\epsilon$ with a
S\'ersic profile is given by
\begin{equation}
  L_{\mathrm tot} \;=\; (1 - \epsilon) \: \frac{2\pi n e^{b_n}}{b_{n}^{2n}} 
  \Gamma(2n) \: I_{e} r_{e}^{2},
\end{equation}
with $\Gamma$ being the gamma function; $r_{e}$ is the half-light
radius along the major axis.  For an exponential disk with observed
ellipticity $\epsilon$, observed central surface brightness $I_{0}$,
and scale length $h$, the luminosity within $R$ is
\begin{equation}
  L(R) \: = \:  (1 - \epsilon)^{C} \: 2 \pi I_{0} h^{2} e^{-R/h}
\end{equation}
and the total luminosity is
\begin{equation}
  L_{\mathrm tot} \;=\; (1 - \epsilon)^{C} \: 2 \pi I_{0} h^{2}.
\end{equation}

The factor $C$ is a correction for disk optical thickness, such that
the corrected, face-on central surface brightness $I_{0,i=0} = (1 -
\epsilon)^{C} I_{0}$; $C = 1.0$ for a transparent disk.  We use $C =
0.61$, the $I$-band value from \citet{tully97}.\footnote{This is the
value for an \textit{outer} disk; it may not be appropriate for inner
disks.} The resulting luminosities and luminosity ratios are listed in
Table~\ref{tab:ratios}.

In Table~\ref{tab:ratios} we also list mass estimates for the bulges
and inner disks.  These are based on mean $\bv$ and $\vi$ colors
obtained from the PC images.  For NGC~3945, this is relatively
straightforward, as there is little dust outside the inner few arc
seconds; we assume the bulge is approximately the same color as the
inner disk ($\bv \approx 0.94$, $\vi \approx 1.22$, corrected for
Galactic extinction).  The situation is more complicated for NGC~2787,
due to the tilted dust lanes.  By making measurements in the most
dust-free region in the outer part of the inner disk, we obtain $\bv
\approx 0.93$ and $\vi \approx 1.15$.  By comparing these values with
the results of \citet{vazdekis96}, assuming a single-burst population,
metallicities near solar, and two possible IMFs (Salpeter and
bimodal), we estimate $V$-band M/L ratios of $\sim 2$ for NGC~2787 and
$\sim 3$ for NGC~3945.

In both cases, the bulges are less luminous than the inner disks ---
for NGC~3945, the bulge is only one-tenth of the inner-disk luminosity
(conversely, NGC~3945's inner disk is actually more luminous than the
whole of NGC~2787!).  Consequently, the bulge-to-total (B/T)
luminosity ratios of these galaxies are small: 0.11 for NGC~2787 and
0.04 for NGC~3945.  As we discuss below, these values are rather low
(unusually so for NGC~3945) when compared with typical S0 values from
the literature.

If we imagine for the moment that the bulges and inner disks really
are just one central, ``bulge'' system, then we gain some insight into
the classification of these galaxies --- and possibly the question of
B/D ratios in general.  For both galaxies, the (inner-disk +
bulge)/total ratio is about 0.4, and corresponding ``B/D'' ratio is
about 0.67.  This is at the low end of B/D ratios for S0 galaxies
(\nocite{graham01-bulges}Graham 2001a, using data from
\nocite{simien86}Simien \& de Vaucouleurs 1986), but not at all
unusual.  The \textit{true} ``B/D'' (that is, bulge/(total $-$ bulge))
and B/T ratios for NGC~2787 are 0.12 and 0.11.  These are below the
known limits for S0 galaxies, and at the low end of Sa--Sab galaxies;
they are more typical of Sab--Sb.  For NGC~3945, the true ``B/D'' and
B/T ratios are both $\approx 0.04$, which are really typical of Sbc
and later galaxies.  We can see that both galaxies are plausibly S0 in
terms of their B/T ratios \textit{if} the inner disk and bulge are
lumped together as the bulge; but if the genuine (round, kinematically
hot) bulge is considered, they are \textit{not} typical S0 galaxies at
all.

%% 
 % (Also true when using Alister's $B$-band B/D ratios, though those are
 % seriously incomplete for Sab and earlier; NGC 2787 appears to be
 % typical for Sab, NGC 3945 comes across as typical of Sb and later; NGC
 % 2787 fits into ``range'' defined by the two S0a galaxies\ldots).
 %%

So is NGC~3945 really a severely misclassified Sc galaxy, and NGC~2787
perhaps better described as Sab?  It is worth remembering that the
Hubble classifications are based on more than (apparent) bulge/disk
ratio.  Both galaxies have no visible spiral arms and no signs of star
formation; NGC~3945 has almost no dust, while the dust in NGC~2787 is
probably the result of an encounter and is not part of its disk(s). 
So both galaxies are clearly S0 in terms of their (outer) disk
structure, which is at least partly why \citet{buta94} listed them as
representative of Hubble type S$0^{+}$.  The implication, then, is not
so much that these galaxies are misclassified, as that the S0 class is
much more heterogeneous in terms of B/T ratios than is commonly
assumed.

\subsection{The Dangers of Global Bulge-Disk Decomposition}

Numerous studies over the years have emphasized measuring bulge and
disk parameters in disk galaxies.  Typically, this is done by
decomposing profiles, either major-axis or azimuthally averaged, into
bulge + disk components.  More recently, this has been expanded to
include two-dimensional fitting
\citep[e.g.,][]{dejong96,moriondo98,khosroshahi00}.  All of these
approaches, however, are predicated on the assumption that the
galaxies can be analyzed into just two components: bulges residing
inside large disks.\footnote{Occasionally, an attempt is made to
account for the presence of bars \citep[e.g.,][]{dejong96,prieto01}.}

\citet{kormendy93} pointed out the possible dangers of relying on such
simplistic, broad-brush decompositions, particularly given the
evidence for the ``bulges'' of some galaxies being kinematically
disk-like (he noted that a significant number of \nocite{kent85}Kent's
[1985,1987] bulge-disk decompositions required highly flattened bulge
components).  One such danger is that true bulge bulge-to-total light
ratios can be significantly overestimated.  As we show below, the
inner disks we find in NGC~2787 and NGC~3945 are large enough and
bright enough to masquerade as bulges in standard bulge-disk
decompositions, resulting in systematic overestimates of the bulge
sizes for these galaxies in the past.  They are thus excellent
examples of how standard bulge-disk decompositions can produce
dramatically misleading results.

Several estimates of bulge size from the literature for NGC~2787 and
NGC~3945 are collected and compared with our measurements in
Table~\ref{tab:bulge-sizes}.  The visual size estimates from
\citet{athan80} are clearly measures of the inner disks.  The $V$-band
major-axis decompositions by \citet{bba98}, which used an $R^{1/4}$
bulge and an exponential disk with an optional hole in the middle, are
clearly poor.  For NGC~3945, this is due to the fact that their
``disk'' fit is to the outer ring, and they warn that such fits are
probably not to be trusted.  For NGC~2787, on the other hand, the
transition from outer disk to lens region, along the major axis,
forces them to introduce a hole to the disk at $r = 44.4\arcsec$;
thus, light belonging to the lens is perforce added, along with the
inner disk, into the ``bulge.''

\begin{figure}
\begin{center}
%% first is whole-page version; second is two-column
%\includegraphics[scale=0.9]{naive_fits}
%\epsscale{0.8}
%\plotone{f14.eps}
\includegraphics[scale=0.52]{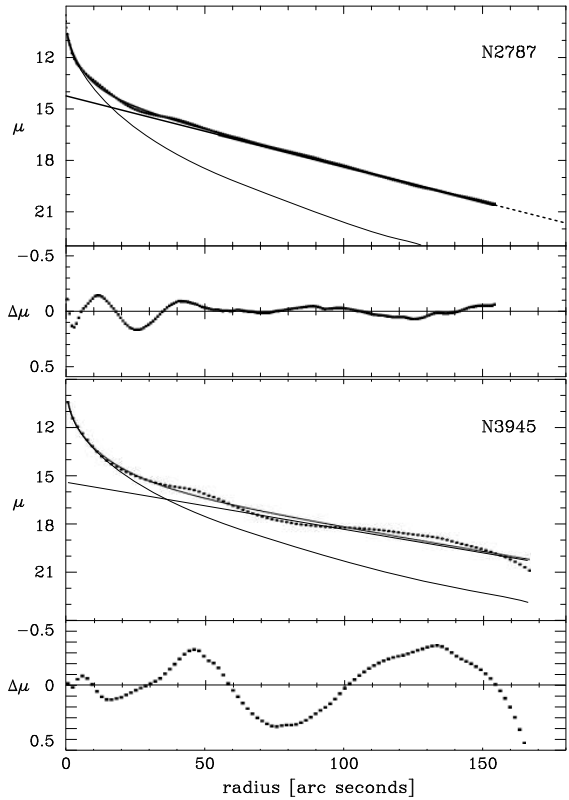}
\end{center}

\caption{``Na\"{\i}ve'' bulge-disk decomposition of the \textit{entire}
azimuthally averaged $R$-band profiles of NGC~2787 (top) and NGC~3945
(bottom), using a S\'ersic profile + exponential (the magnitude scale is
arbitrary for both galaxies).  While there are clear problems with the
fit for NGC~3945, owing to the absence of a traditional outer disk,
the fit for NGC~2787 looks quite good.  In both cases, the inner disks
masquerade as large ``bulges.''\label{fig:naive}}

\end{figure}

In Figure~\ref{fig:naive} we also show simple bulge-disk
decompositions for the azimuthally averaged $R$-band profiles of the
\textit{whole} galaxies.  Azimuthally averaged profiles eliminate some
of the difficulties associated with major-axis profiles, since the
problems created by relative bar orientation (does the profile run
close to the bar major axis?  does it cross a lens region?)  are
lessened.  These fits were done with the method outlined above in
Section~\ref{sec:decomp}, including seeing correction using the median
FWHM of stars in each image and the use of S\'ersic profiles for the
bulges.

The fit for NGC~3945 is quite poor, because there simply is no clear
outer, single-exponential disk in this galaxy \citep[such
non-exponential profiles are typical of barred galaxies with prominent
outer rings; e.g.,][]{dev75,buta01}.  One could try alternate
approaches, such as using the inner part of the major-axis profile,
perhaps treating the lens region as an underlying exponential
component --- note that it is clearly \textit{not} flat, as is often
assumed for lenses, and in fact has \textit{two} slopes.  Nonetheless,
one might still be tempted to suppose that an approximate bulge
profile has been derived, even if the disk fit is not accurate. 
Taking the fits at face value, and assuming $\epsilon = 0.3$ and 0.35
for the ``bulge'' and ``outer disk,'' respectively, yields an $R$-band
B/T ratio of 0.6 (not too different from the $B$-band value of 0.4
which \nocite{kormendy-ill83}Kormendy \& Illingworth 1983 estimated
``from the morphology'').

% (Alister gets: $r_e = 12.9\arcsec$, $n = 2.51$, $\mu_{e} = 13.853$; 
% ``disk'': $h = 37.1\arcsec$, $\mu_{0} = 15.433$).  Resulting B/T 
% ratio = 0.62  and $r_{e}/h = 0.35$
% 
% Note that using the true bulge fit, we get $r_{e}/h = 0.031$;
% this is at the very lower end of what MacArther et al 02 find,
% and seeming only plausible for Sb or later.

The fit for NGC~2787 is better, though it can be seen that the
azimuthally averaged bar region ($a \sim 30$--50\arcsec) produces some
deviations, and a single S\'ersic profile is not very good at fitting
the combination of bulge + inner disk; the latter phenomenon could
serve as a useful warning.  Nonetheless, it is still tempting to
assume that one can (approximately) measure the bulge this way.  The
supposed $R$-band B/T ratio would then be 0.4 (assuming $\epsilon =
0.3$ and 0.4 for the ``bulge'' and outer disk, respectively).  This
is, in fact, about what we get if we add our true bulge and inner-disk
fits together and compare them to the total luminosity in the $I$
band, so the ``na\"{\i}ve'' fit does indeed lump the true bulge and
inner disk together into one larger, composite ``bulge.''

% (Alister gets: $r_e = 10.57\arcsec$, $n = 2.35$, $\mu_{e} = 14.02$; 
% ``disk'': $h = 26.35\arcsec$, $\mu_{0} = 14.25$).  Resulting B/T 
% ratio = 0.4, with $r_{e}/h = 0.40$
% 
% Note that using the true bulge fit, we get $r_{e}/h = 0.16$; this
% is somewhat low, but plausible for any of MacArthur et al's range
% of Hubble type (basically Sab and later).

In both cases, the na\"{\i}ve interpretation is that we can identify the
bulge regions and the parameters of the bulges from these fits.  Given
our analysis in the preceding sections, these parameters would be
rather wrong: we would overestimate the bulge effective radius by
factors of two (in the case of NGC~2787) to ten (for NGC~3945);
similarly, the bulge luminosities would be overestimated by factors of
two to $\sim 30$!

\subsection{Frequency of Bright Inner Disks and Bulge Overestimation}

How common is it that a significant fraction of a galaxy's central
light is due to something like an inner disk, rather than to the
bulge?  In \nocite{erwin-sparke02}Erwin \& Sparke's (2002) sample of
barred S0--Sa galaxies, about one-third of the twenty S0 galaxies had
potential inner disks.  As they pointed out, \textit{some} of these
are probably inner bars or unresolved nuclear rings; moreover, we have
been studying here the two largest inner disks from that sample, so we
are probably sampling the upper end of the distribution --- other
inner disks may be less luminous and a lower fraction of the central
light.  From this, we can argue that $\sim 10$\% of (barred) S0
galaxies may have ``composite bulges,'' where the true bulge is at
most half the luminosity of the central subsystem.  But this is
probably a lower limit, since Erwin \& Sparke relied on ellipse fits
to identify possible inner disks.  In galaxies which are close to
face-on, an inner disk will not be distinguishable from a pure bulge
using ellipse fits.  If we sort all the galaxies of their sample by
inclination, then the median inclination of the upper half (higher
inclinations) is 52\arcdeg, versus 33\arcdeg{} for the less inclined
half.  The median inclination of the inner-disk galaxies is 52\arcdeg;
a Kolmogorov-Smirnov test confirms that the inner-disk galaxies are
indeed drawn from the upper half of the inclinations (probability =
81\%, versus a 0.04\% probability that the inner-disk galaxy
inclinations are consistent with being drawn from the \textit{lower}
half of the inclinations).  This suggests, very approximately, that
Erwin \& Sparke detected only half of the inner disks in their sample,
in which case the frequency of large inner disks is closer to $\sim
20$\% of barred S0's.

In a study of edge-on S0 galaxies, \citet{seifert96} found inner disks
in about \textit{half} of their fifteen galaxies.  The problem here is
that, as we noted in the Introduction, while we know these structures
are flat, it is unclear how many are \textit{disks}, and not bars or
rings; thus, their detection rate is probably best seen as an
\textit{upper} limit.  (We also do not know how many of their galaxies
were barred, which makes comparison with the barred-galaxy sample of
Erwin \& Sparke a little difficult.)  Assuming that most of these are
disks, we can ask how many of them are comparable in size to those of
NGC~2787 and NGC~3945, since Seifert \& Scorza provide approximate
scale lengths for the inner disks.  Defining $f_{25}$ as the ratio of
scale length $h$ to $R_{25}$, we find $f_{25} = 0.099$ for NGC~2787
and 0.035 for NGC~3945.  Three of Seifert \& Scorza's inner disks have
similar sizes (the rest are smaller): NGC~2549 ($h \approx 5\arcsec
\sim 280$ pc and $f_{25} \approx 0.043$); NGC~4026 ($h \approx
5\arcsec \sim 380$ pc and $f_{25} \approx 0.032$); and NGC~7332 ($h
\approx 5\arcsec \sim 560$ pc and $f_{25} \approx 0.041$).\footnote{For
NGC~2549 and NGC~7332, we used SBF distances from \citet{tonry01}; for
NGC~4026, we used $V_{\rm vir} = 1167$ from LEDA and $H_{0} = 75$
\kms{} Mpc$^{-1}$.}  This implies that $\lesssim 20$\% of S0's harbor
large inner disks.

Taken together, these results suggest that the fraction of S0's with
significantly composite bulges is $\sim 20$\%.  But this is not the
only way for a galaxy to have a composite bulge.  About a third of
Erwin \& Sparke's barred S0's had inner \textit{bars}.  While some of
these were quite weak features, two of them (NGC~2859 and NGC~2950)
produced severe distortions in the inner isophotes, such that
\citet{kormendy82a} referred to both as having ``triaxial'' bulges;
moreover, he found that both galaxies had $\vmsstar \approx 1.2$,
suggesting the dominance of disk-like over bulgelike kinematics in the
inner-bar region.  It is plausible to imagine that in these galaxies,
too, the true bulge contribution to the luminosity is significantly
less than what might be estimated by casual inspection, or by a global
bulge-disk decomposition.  So the composite-bulge fraction (galaxies
with large inner disks plus those with large inner bars) could be as
high as $\sim 30$\% in barred S0's.

It is curious that Erwin \& Sparke detected inner disks almost
exclusively in S0 galaxies, even though about half their sample were
S0/a or Sa and inner \textit{bars} were found with roughly equal
frequency in all three Hubble types.  This suggests that inner disks
might be largely confined to S0 galaxies, though why this should be so
is not at all clear.  There are presumably \textit{some} sizeable
inner disks in spiral galaxies; \citet{kormendy93} suggested that the
fraction of inner-disk--dominated ``bulges'' should increase to later
Hubble types.  The Sa galaxy NGC~4594 (the Sombrero Galaxy) has an
inner disk with scale length $\approx 170$ pc, and \citet{pizzella02}
found an $h \sim 250$ pc inner disk in the Sb NGC~1425; in both cases,
however, $f_{25} \lesssim 0.015$, so they are still rather
small.\footnote{Using the scale length from \citet{emsellem94} and the
SBF distance from \citet{tonry01} for NGC~4594; Cepheid distance for
NGC~1425 from \citet{freedman01}.} An even better example may be
NGC~3368 (M96), an SABab galaxy with both an inner bar and a possible
inner disk visible in near-IR images.  Although an exponential scale
length is not available, the maximum-ellipticity radius of the inner
disk is $\sim 1$ kpc --- almost identical to that of NGC~3945. 
\citep[for details, see][]{erwin03-db}.  The evidence for a
bar-inside-disk-inside-bar structure makes this galaxy an even closer
analog to NGC~3945.  Other non-S0 examples may include NGC~1433 (SBab)
and NGC~6300 (SBb), which \citet{buta01} found to have highly
flattened isophotes in the outer parts of their ``bulges regions.''

\subsection{Origins and Dynamics} %[x]
\label{sec:origins}

We can speculate briefly on the origin of these inner disks.  The most
plausible mechanism is bar-driven gas inflow, resulting in a central
accumulation of gas inside the bar.  Once the gas density is
sufficiently high, star formation is a likely consequence.  This would
be a clear example of bar-driven secular evolution \citep[e.g.,][and
references therein]{kormendy93}.  Is this consistent with ideas about
secular evolution along the Hubble sequence, especially those where it
occurs due to bars \citep[e.g,][]{hasan93,norman96}?

In once sense, the answer is yes.  As we have pointed out above, the
presence of inner disks such as these can contribute to classifying
their galaxies as early-type, since the inner disks masquerade as
bulges (``pseudobulges'' in Kormendy's terminology).  On the other
hand, the inner disks are \textit{not} classical bulges --- they are
disks.  Thus, models where bars \textit{dissolve} into spheroidal,
kinematically hot bulges \nocite{norman96}(e.g., Norman et al.\ 1996)
do not work here --- not the least because these galaxies still have
bars!  Both NGC~2787 and NGC~3945 do appear to have small, classical
bulges embedded within their inner disks; we do not know if these
bulges were formed before, during, or after the inner disk formation,
though this is clearly an interesting question.

The absence of a clear color difference between the inner disks and
the surrounding stars \citep{erwin99,erwin-sparke03} suggests that
star formation ceased some time ago; the presence of off-plane gas in
NGC~2787 also argues against a significant, recent in-plane
accumulation of gas in the inner-disk region.  Is there evidence for
present-day inner-disk formation in nearby galaxies?  \textit{Current}
circumnuclear star formation in barred galaxies is a well-studied
phenomenon, though it usually takes place in a fairly narrow ring, not
a disk.  One intriguing example is the double-barred Sa galaxy
NGC~4314, where there is active star formation in a circumnuclear ring
(surrounding the inner bar) with diameter $\ap 10\arcsec$ and evidence
for a recent but older epoch of star formation outside, in a zone of
diameter 20--25\arcsec{} possibly associated with $x_{2}$ orbits in
between two ILRs \citep{benedict02}.  As Benedict et al.\ suggest,
this may indicate a process where the accumulated gas shrinks in
radius over time, with star formation thus proceeding outside-in
\citep{combes92}.  (An alternate scenario has star formation mainly in
the ring, with the dynamical influence of the inner bar driving older
star clusters outward; \nocite{kenney99}Kenney, Friedli, \& Benedict
1999.)  Another possibly relevant galaxy is NGC~1317, where there are
\textit{two} star-forming nuclear rings inside an outer bar: an
elliptical ring surrounding an inner bar, and a circular ring further
out \citep{crocker96}.  Since NGC~3945 is also double-barred, with a
(stellar) nuclear ring surrounding the inner bar, it is tempting to
identify NGC~1317 and NGC~4314 as precursor objects, where inner disks
may be forming as we watch.  Whether such outside-in or double-ring
circumnuclear star formation can actually produce the massive,
exponential inner disks we see in NGC~2787 and NGC~3945 is unknown.

However they might form, how do such disks \textit{survive} inside
bars?  Is it dynamically plausible for a massive disk to exist inside
a strong bar?  We have suggested, based on the presence and tentative
location of ILRs in both galaxies, that the inner disks may be
supported by $x_{2}$ orbits within the bars.  (The actual radial range
spanned by the $x_{2}$ orbits cannot be accurately determined from our
crude ILR identifications, but must be found by orbit integrations in
a realistic potential.)  A possible objection to this idea is the fact
that $x_{2}$ orbits are elongated perpendicular to bars (indeed, they
have been sometimes suggested as the basis for perpendicular inner
bars), while we have been arguing that the inner disks are essentially
circular.  An alternate scenario is suggested by the findings of
\citet{heller96}, who studied the effects of massive circular and
elliptical nuclear rings on the orbits within bars.  Intriguingly, the
presence of a circular ring tended to make the bar-supporting $x_{1}$
orbits --- normally elongated parallel with the bar ---
\textit{rounder} in the vicinity of the ring (see their Section~4.2
and Fig.~9).  It could be that the presence of a disk would cause a
similar phenomenon, and in this case the circularized, inner $x_{1}$
orbits would be a plausible mechanism for supporting the inner disks.

\section{Summary}
\label{sec:summary}

We have presented a morphological, photometric, and kinematic
analysis, using both ground-based and \textit{HST} images and
ground-based long-slit spectra, of two barred S0 galaxies with
distinct, kiloparsec-scale \textit{inner disks} inside their bars.  We
show that these disks are probably flat and circular, and are thus
geometrically disk-like.  They also have clear exponential profiles,
making them similar to classical large disks as well.

Published and newly-reduced archival stellar kinematics from long-slit
spectra indicate that the inner disks are dominated by rotation rather
than by random stellar motions, so they are probably kinematically
disk-like \citep[as previously pointed out for one of the galaxies
by][]{kormendy82a}.  We also find evidence for inner Lindblad
resonances in the inner-disk region of both galaxies.  This suggests
that the inner disks could be supported by $x_{2}$ (bar-perpendicular)
orbits in the bar potential; alternately, the influence of a massive
inner disk on the potential may make the bar-supporting $x_{1}$ orbits
more nearly circular in the inner-disk region, and thus capable of
supporting the disk.

Due to their short scale lengths and high central surface brightness,
these inner disks appear as central excesses above the outer-disk
luminosity profile.  Consequently, they can be --- and indeed have
been --- erroneously classified as the ``bulges'' of these galaxies,
and are thus good examples of what \citet{kormendy01} and
\citet{kormendy-texas} term ``pseudobulges'' \citep[see 
also][]{kormendy93}.

It is important to note that both galaxies \textit{do} in fact have
central bulges which are distinct from the inner disks.  These ``true
bulges'' appear as excesses above the inner disk profile, are rounder
in projection than the inner disks (suggesting they are oblate or
mildly triaxial spheroids), and have non-exponential ($\approx
r^{1/2}$) light profiles.  The true bulges constitute much smaller
fractions of the total galaxy light than a simplistic bulge-disk
decomposition of the global light profiles would indicate: such
decompositions assign light from the inner disk to the bulge
component.  In NGC~2787, the inner disk has twice the luminosity of
the true bulge; in NGC~3945, the inner disk is almost ten times more
luminous than the true bulge, making the true bulge/total ratio of the
galaxy $\approx $5\%, a value more typical of Sc galaxies than S0
galaxies.  We estimate that $\sim 20$--30\% of S0 galaxies may have
similar ``composite bulges'' or ``pseudobulges,'' where a significant
fraction of the inner light comes from an inner disk or bar, rather
than a spheroidal, kinematically hot bulge.  This means that a
substantial number of bulge sizes, luminosities, and bulge-to-disk
ratios may have been overestimated.

\acknowledgments

We would like to thank Enrico Corsini for supplying us with the
stellar and gas kinematics for NGC~3945 \citep[originally published in
][]{bertola95} and Lodovico Cocatto for the unpublished gas kinematics
of NGC~2787.  We also thank the referee, John Kormendy, for careful
reading and useful comments.

Based on observations made with the Isaac Newton Group of Telescopes
operated on behalf of the UK Particle Physics and Astronomy Research
Council (PPARC) and the Nederlanse Organisatie voor Wetenschappelijk
Onderzoek (NWO) on the island of Tenerife in the Spanish Observatorio
del Roque de Los Muchachos of the Instituto de Astrofisica de
Canarias.  This research was partially based on data from the ING
archive, and is also based on observations made with the NASA/ESA
\textit{Hubble Space Telescope}, obtained from the data archive at the
Space Telescope Institute.  STScI is operated by the association of
Universities for Research in Astronomy, Inc.\ under the NASA contract
NAS 5-26555.

This research has made use of the NASA/IPAC Extragalactic Database
(NED) which is operated by the Jet Propulsion Laboratory, California
Institute of Technology, under contract with the National Aeronautics
and Space Administration.  We also made use of the Lyon-Meudon
Extragalactic Database (LEDA; http://leda.univ-lyon1.fr).  Finally, 
this research was partly supported by grant AYA2001-0435 of the 
Spanish Ministerio de Ciencia y Tecnolog\'{\i}a.

\appendix

% References:

% Tables:
\clearpage
\begin{deluxetable}{lrrrr}
\tablewidth{0pt}
\tablecaption{General Characteristics of NGC~2787 and NGC~3945\label{tab:general}}
\tablecolumns{5}
\tablehead{\colhead{Parameter} & \colhead{NGC~2787} & 
\colhead{Source} & \colhead{NGC~3945} & \colhead{Source}}
\startdata
Morphological Type          & SB(r)$0^{+}$ &  RC3  & (R)SB(rs)$0^{+}$ & RC3 \\
$R_{25}$                    &  95\arcsec{} &  RC3  & 157\arcsec{}     & RC3 \\
Heliocentric $V_{\rm rad}$  &  696 \kms    &  LEDA & 1233 \kms        & LEDA \\
Distance                    & 7.5 Mpc      &  1    &  19.8 Mpc        & LEDA \\
Linear scale (1\arcsec)     &  36 pc       &       &  96 pc           & \\
Major axis PA               & 109\arcdeg   &  2    & 158\arcdeg       & 3 \\
Inclination                 &  55\arcdeg   &  2    & $\sim50\arcdeg$  & 3 \\
$B_{T}$                     &  11.27       &  LEDA &  11.54           & LEDA \\
$V_{T}$                     &  10.37       &  LEDA &  10.65           & LEDA \\
$I_{T}$                     &   8.80       &  LEDA & \nodata          & LEDA \\
$M_{B}$                     & -18.10       &       & -19.94           & \\
$M_{V}$                     & -19.00       &       & -20.83           & \\
$M_{I}$                     & -20.57       &       & \nodata          & \\
Galactic extinction $A_{B}$ &   0.57       &  4    &   0.12           & 4 \\
\enddata

\tablecomments{Unless otherwise noted, measurements are from this
paper.  The distance to NGC~3945 uses radial velocity, corrected for
Virgo-centric infall, from LEDA and $H_{0} = 75$ \kms{} Mpc$^{-1}$. 
Absolute magnitudes use total magnitude from LEDA and the distances
listed here.  Other sources: 1 = Tonry et al.\ 2001; 2 = Erwin \&
Sparke 2002b; 3 = Erwin \& Sparke 1999; 4 = Schlegel, Finkbeiner, \&
Davis 1998, via NED; LEDA = Lyon-Meudon Extragalactic Database.}

\end{deluxetable}

\begin{deluxetable}{lr}
\tablewidth{0pt}
\tablecaption{Instrumental Setup for Spectroscopic Observations of 
NGC~2787\label{tab:setup}} 
\tablecolumns{2}
\tablehead{\colhead{Parameter} & \colhead{Value}}
\startdata
Date                                           & 24--26 Dec 1995 \\ 
Spectrograph                                   & ISIS           \\  	 
Grating ($\rm lines\;mm^{-1}$)                 & 1200           \\ 
Detector                                       & TK1024         \\
Pixel size ($\rm \mu m$)                       & $24\times24$   \\
Pixel binning                                  & $1\times1$     \\  
Scale ($\rm ''\;pixel^{-1}$)                   & 0.36           \\ 	   
Reciprocal dispersion (\AA{} pixel$^{-1}$)     & 0.40           \\ 
Slit width ($''$)                              & 1.3            \\
Slit length ($'$)                              & 4.0            \\
Spectral range (\AA)                           & 5000--5400     \\
Comparison lamp                                & CuNe+CuAr      \\  
Instrumental FWHM (\AA)                        & $0.997\pm0.119$\\ 
Instrumental $\sigma$ (\kms)                   & 25             \\ 
Seeing FWHM ($''$)                             & 1.3--1.8       \\ 
Exposure: major axis (PA $= 117\arcdeg$)       & 2 x 600s       \\
Exposure: offset 23\arcsec{} NE                & 2 x 1800s      \\
Exposure: offset 23\arcsec{} SW                & 1 x 1800s      \\
\tablecomments{The instrumental $\sigma$ was measured at \oiii.}
\enddata
\end{deluxetable}  

\clearpage

%% SUBMITTED VERSION: use the next three lines and delete the
%% following two ``mini-tables''
%\input{tab3.tex}
%\clearpage
%\input{tab4.tex}

%% PREPRINT:
% Stellar kinematics for NGC 2787 (online) --- DELETE FOR
% SUBMITTED VERSION
%tab:stellar-kin
%% 
\begin{deluxetable}{llllll}
\tablewidth{0pt}
\tablecaption{NGC 2787: Major-axis Kinematics\label{tab:ma-kin}} 
\tablecolumns{6}
\tablehead{\colhead{$r_{\star}$} & \colhead{$V_{\star}$} &
\colhead{$\sigma_{\star}$} & \colhead{$r_{g}$} &
\colhead{$V_{g}$} & \colhead{$\sigma_{g}$}}
\startdata
$-60.8$ &   209  & 106 & $-19.3$&   123 &  40  \\
$-36.9$ &   188  & 118 & $-16.6$&   122 &	8  \\
$-22.5$ &   158  & 130 & $-14.8$&   136 &  22  \\
$-16.3$ &   150  & 151 & $-13.0$&   132 &	0  \\
$-12.8$ &   132  & 182 & $-11.2$&   123 &  44  \\
$-10.3$ &   116  & 182 &  $-9.9$&   124 &  34  \\
$-8.4$  &   106  & 177 &  $-9.2$&   113 &  34  \\
$-6.7$  &   117  & 178 &  $-8.5$&   120 &  26  \\
$-5.5$  &   114  & 218 &  $-7.7$&   103 &  31  \\
$-4.4$  &   110  & 220 &  $-7.0$&   103 &  20  \\
$-3.3$  &   100  & 241 &  $-6.3$&   115 &  25  \\
$-2.4$  &    69  & 227 &  $-5.6$&    82 &  38  \\
$-1.7$  &    65  & 241 &  $-4.7$&    90 &  26  \\
$-1.2$  &    63  & 215 &  $-4.3$&    91 &  46  \\
$-0.8$  &    44  & 228 &  $-4.0$&    98 &  44  \\
$-0.4$  &    22  & 246 &  $-3.6$&    98 &  27  \\
$-0.1$  &  $-12$ & 257 &  $-3.2$&    96 &  36  \\
0.3     &  $-23$ & 257 &  $-2.9$&    96 &  53  \\
0.6     &  $-51$ & 222 &  $-2.5$&    91 &  32  \\
1.0     &  $-73$ & 200 &  $-2.2$&    99 &  41  \\
1.4     &  $-60$ & 214 &  $-1.8$&    92 &  45  \\
1.9     &  $-76$ & 242 &  $-1.4$&    87 &  62  \\
2.6     & $-112$ & 189 &  $-1.1$&    89 &  76  \\
3.3     & $-100$ & 192 &  $-0.7$&    85 &  85  \\
4.2     & $-121$ & 179 &  $-0.4$&    73 & 109  \\
5.3     &  $-83$ & 196 &    0.0 &    41 & 111  \\
6.5     & $-105$ & 180 &    0.4 &    10 & 107  \\
8.2     & $-110$ & 169 &    0.7 &  $-14$&  91  \\
10.1    & $-130$ & 175 &    1.1 &  $-47$&  60  \\
12.6    & $-119$ & 158 &    1.4 &  $-56$&  57  \\
16.1    & $-136$ & 135 &    1.8 &  $-63$&  51  \\
22.3    & $-127$ & 168 &    2.2 &  $-63$&  50  \\
37.4    & $-217$ &  73 &    2.5 &  $-67$&  26  \\
60.6    & $-155$ & 144 &    2.9 &  $-66$&  26  \\
        &        &     &    3.2 &  $-82$&  34  \\
        &        &     &    3.6 &  $-85$&  41  \\
        &        &     &    4.0 &  $-84$&  50  \\
        &        &     &    4.3 & $-104$&  14  \\
        &        &     &    4.7 & $-102$&  25  \\
        &        &     &    5.0 &  $-91$&  26  \\
        &        &     &    5.4 &  $-84$&  31  \\
        &        &     &    5.8 &  $-84$&  25  \\
        &        &     &    6.3 &  $-88$&  37  \\
        &        &     &    7.0 &  $-96$&  39  \\
        &        &     &    7.7 &  $-83$&  47  \\
        &        &     &    8.5 &  $-53$&  52  \\
        &        &     &    9.2 &  $-88$&  48  \\
        &        &     &    9.9 &  $-83$&  25  \\
        &        &     &   10.8 & $-110$&  38  \\
        &        &     &   12.2 & $-101$&  36  \\
        &        &     &   14.0 & $-112$&  25  \\
        &        &     &   15.8 & $-123$&  16  \\
        &        &     &   17.6 & $-115$&  23  \\
\tablecomments{Stellar and gaseous kinematics along the major axis
(actually PA $= 117\arcdeg$) of NGC~2787.  $r_{\star}$, $V_{\star}$,
and $\sigma_{\star}$ are the radius, velocity, and velocity dispersion
of the stars; the ``g'' subscript refers in a similar manner to the
ionized-gas kinematics.}
\enddata
\end{deluxetable}

%% PREPRINT:
% Offset kinematics for NGC 2787 (online) -- DELETE FOR SUBMITTED
% VERSION
%tab:offset-kin
\begin{deluxetable}{rrrrrr}
\tablewidth{0pt}
\tablecaption{NGC 2787: Offset Stellar Kinematics\label{tab:offset-kin}} 
\tablecolumns{6}
\tablehead
{
\multicolumn{3}{c}{$23\arcsec$ NE offset} &
\multicolumn{3}{c}{$23\arcsec$ SW offset}\\
\colhead{$r_{\star}$ ($\arcsec$)} & \colhead{$V_{\star}$ (\kms)} &
\colhead{$\sigma_{\star}$ (\kms)} & \colhead{$r_{\star}$ ($\arcsec$)} &
\colhead{$V_{\star}$ (\kms)} & \colhead{$\sigma_{\star}$ (\kms)}
}
\startdata

$-56.5$ &   185  & 123   & $-57.1$ &   170  &   48  \\
$-47.2$ &   165  &  68   & $-46.9$ &   163  &  118  \\
$-36.9$ &   153  &  69   & $-35.9$ &   122  &  124  \\
$-28.8$ &   143  &  84   & $-28.6$ &   110  &  178  \\
$-22.0$ &   105  & 100   & $-23.4$ &    81  &  106  \\
$-15.9$ &   115  & 102   & $-19.5$ &    53  &  119  \\
$-10.7$ &    56  &  99   & $-16.1$ &    48  &  107  \\
 $-6.2$ &    49  & 114   & $-12.9$ &    56  &  151  \\
 $-2.1$ &    50  & 136   & $ -9.5$ &    22  &  115  \\
   1.7  &    11  & 100   & $ -5.9$ &    10  &  146  \\
   5.3  &   $-8$ &  93   & $ -2.1$ &  $-16$ &  112  \\
   8.9  &  $-22$ & 105   &    2.1  &  $-17$ &  112  \\
  12.3  &  $-38$ &  86   &    6.4  &  $-59$ &  160  \\
  15.7  &  $-55$ &  94   &   10.8  &  $-48$ &  161  \\
  19.5  &  $-61$ &  66   &   15.8  &  $-52$ &  134  \\
  23.9  &  $-91$ & 119   &   21.8  &  $-98$ &  164  \\
  30.1  &  $-91$ &  97   &   28.8  & $-148$ &  105  \\
  39.7  & $-119$ &  62   &   37.8  & $-178$ &  135  \\
  52.0  & $-145$ &  74   &   50.8  & $-156$ &  173  \\

\tablecomments{Stellar kinematics for NGC~2787 along offset slit
positions (PA $= 117\arcdeg$, offset the specified distances from the
galaxy nucleus).  $r_{\star}$, $V_{\star}$, and $\sigma_{\star}$ are
the radius along the slit, velocity, and velocity dispersion of the
stars; for each slit position, $r_{\star} = 0$ marks the closest point
to the nucleus.}

\enddata
\end{deluxetable}

\clearpage

% Bulge-disk decompositions:
\begin{deluxetable}{lcrrrrr}
\tabletypesize{\footnotesize}
\tablewidth{0pt}
\tablecaption{Bulge-Disk Decompositions for Inner 
Disks\label{tab:decomp}}
\tablecolumns{7}
\tablehead
{
\colhead{Galaxy} & \colhead{Filter} & \multicolumn{2}{c}{Inner Disk} &
\multicolumn{3}{c}{Bulge} \\
\colhead{} & \colhead{} & \colhead{$\mu_{0}$} & \colhead{$h$ (\arcsec/pc)} &
\colhead{$\mu_{e}$} & \colhead{$r_{e}$ (\arcsec/pc)} & \colhead{$n$}
}
\startdata
%  Name  filter     mu_{0}    h             mu_{e}     r_{e}      n
NGC 2787 &  $I$ &   16.20 &  8.76/318 &     17.28  &   4.27/155  &  2.31 \\
%(outer disk) & $I$ & xx.xx & x.xx &     ---    &   ---   &  ---  \\
\\
NGC 3945 &  $I$ &   15.71 &  5.54/532 &     16.03  &   1.15/110  &  1.85 \\
\enddata

\tablecomments{For each galaxy, we give the results of a bulge-disk
decomposition applied to the inner-disk region.  Fits are to
major-axis profiles of F814W WFPC2 images, converted to $I$ as
described in the text, with corrections for Galactic extinction.}

\end{deluxetable}

% Bulge and disk luminosities and ratios:
\begin{deluxetable}{lcrcrrcrr}
\tabletypesize{\footnotesize}
\tablewidth{0pt}
\tablecaption{Bulge and Inner Disk Magnitudes, Masses, and Luminosity Ratios\label{tab:ratios}}
\tablecolumns{7}
\tablehead
{
\colhead{Galaxy} & \colhead{Filter} & $M_{\rm disk}$ & mass ($\Msun$) &
ID/T & $M_{\rm bulge}$ & mass ($\Msun$) & B/T & B/ID
}
\startdata
%  Name   band     M_d   mass              ID/T
%  M_b     mass               B/T      B/ID
NGC 2787 & $I$ & -19.1 & $3\times10^{9}$ & 0.26 & 
  -18.2 & $1\times10^{9}$  &  0.11  &  0.42 \\
%(outer disk) & $I$ & xx.xx & x.xx &     ---    &   ---   &  ---  \\
\\
NGC 3945 & $V$ & -19.8 & $2\times10^{10}$ & 0.37 &  
  -17.4  & $2\times10^{9}$  &  0.04  &  0.12 \\
         & $I$ & -21.0 &  &   ---  &  -18.6  &   &  ---   &  0.12 \\
\enddata

\tablecomments{Absolute bulge and inner-disk (ID) magnitudes and
masses, using fits from Table~\ref{tab:decomp} and distances from
Table~\ref{tab:general}.  Because NGC~3945 lacks a total $I$-band
magnitude, we convert the $I$-band inner-disk and bulge measurements
to the $V$-band, using $\vi = 1.2$ (see text).  For NGC~2787, we
assumed inner-disk $\epsilon = 0.35$ and truncation at $R_{t} =
20\arcsec$; for NGC~3945, $\epsilon = 0.36$ and $R_{t} = 17\arcsec$. 
For both galaxies, bulge $\epsilon = 0.1$.  ID/T = ratio of inner-disk
to total luminosity; B/T = ratio of bulge to total luminosity; B/ID =
ratio of bulge to inner-disk luminosity.  See text for mass
estimation.}

\end{deluxetable}

% Comparison of bulge sizes:
\begin{deluxetable}{lrcrr}
\tabletypesize{\footnotesize}
\tablewidth{0pt}
\tablecaption{Comparison of Bulge Size 
Measurements\label{tab:bulge-sizes}}
\tablecolumns{5}
\tablehead
{
\colhead{Galaxy} & \colhead{Bulge radius (\arcsec)} & 
\colhead{Filter} & \colhead{Decomposition} & \colhead{Source}
}
\startdata
% Name      radius   filter   method          source
NGC 2787  &    4.6  & $I$ & inner-disk        & 1 \\
          &   10.6  & $R$ & full profile      & 1 \\
          &   16.3  & $V$ & full profile (MA) & 2 \\
          &   21.5  & $B$ & visual estimate   & 3 \\
\\
NGC 3945  &    1.2  & $I$ & inner-disk        & 1 \\
          &   12.9  & $R$ & full profile      & 1 \\
          &   16.3  & $V$ & full profile (MA) & 2 \\
          &   16.0  & $B$ & visual estimate   & 3 \\
		  &   17.0  & $B$ & visual estimate   & 4 \\
\enddata

\tablecomments{Comparison of bulge size measurements.  All except for
the ``visual estimate'' measurements are effective radii from
bulge-disk decompositions; the ``full profile'' decompositions use
either the azimuthally averaged profile or a major-axis cut (``MA''). 
Because all methods \textit{except} the inner-disk decompositions
assume that the bulge and (outer) disk are the only components in each
galaxy, they force the inner disk to be part of the bulge and thus
overestimate the bulge size.  Sources: 1 = this study; 2 = Baggett,
Baggett, \& Anderson 1998; 3 = Athanassoula \& Martinet 1980; 4 = 
Kormendy 1982.}

\end{deluxetable}

\clearpage

% Figure Captions:
\clearpage

\onecolumn

% Overall isohpotes; Bar and Disk unsharp masks

% Fancy kinematics+isophotes figure for NGC 2787

% Fancy kinematics+isophotes figure for NGC 3945:

%% PREPRINT:
\twocolumn

%% PREPRINT:
\onecolumn

% Bar profiles:

% Inner-disk profiles:

%% PREPRINT:
\twocolumn

% bulge-disk decompositions:

% Truncation of inner disk in NGC 2787:

%% 
 % \begin{figure}
 % \begin{center}
 % \includegraphics{n3945_colors}
 % \end{center}
 % 
 % \caption{Color profiles for the bulge and inner disk of NGC~3945, 
 % based on the HST WFPC2 images.  Colors have been corrected for 
 % Galactic reddening.  These profiles are based on ellipse fits to the 
 % F814W image.\label{fig:n3945-colors}}
 % 
 % \end{figure}
 %%

\end{document}